\documentclass[fleqn,usenatbib]{mnras}
\usepackage{soul}
\usepackage{graphicx}
\usepackage{amssymb }
\usepackage[T1]{fontenc}
\usepackage{multirow}
\usepackage{url}
\usepackage{hyperref, amsmath}
\usepackage{caption}
\usepackage{subcaption}
\usepackage{siunitx}
\usepackage[normalem]{ulem} 

\begin{document}
\title{Non-Separable Halo Bias from High-Redshift Galaxy Clustering}
\author[Mons et al]{
Emy Mons,$^{1}$\thanks{E-mail: emymons92@gmail.com}
Vipul Prasad Maranchery,$^{2,3,4}$
M. S. Suryan Sivadas$^{1}$
and Charles Jose$^{1}$\thanks{E-mail: charles.jose@cusat.ac.in }
\\
$^{1}$Department of Physics, CUSAT, Cochin, 682022, India\\
$^{2}$Dipartimento di Fisica, Università di Roma Tor Vergata, Via
della Ricerca Scientifica, 1, Roma 00133, Italy\\
$^{3}$Italian Space Agency (ASI), Via del Politecnico, 00133, Roma, Italy\\
$^{4}$Dipartimento di Fisica, Università di Roma "La Sapienza", P.le
A. Moro 5, Roma 00185, Italy\\
}
\newcommand{\cjc}[1]{\textcolor{red}{CJ: #1}}
\newcommand{\emc}[1]{\textcolor{green}{EM: #1}}
\newcommand{\msc}[1]{\textcolor{magenta}{MS: #1}}

\date{Accepted 2026 January 28. Received 2026 January 6; in original form 2025 June 10}
\pubyear{\the\year{}}

\label{firstpage}
\pagerange{\pageref{firstpage}--\pageref{lastpage}}
\maketitle

\begin{abstract}
The halo model provides a powerful framework for interpreting galaxy clustering by linking the spatial distribution of dark matter haloes to the underlying matter distribution. 
A key assumption within the halo bias approximation of the halo model is that, on sufficiently large scales, the halo bias between two halo populations is a separable function of the mass of each population. 
In this work, we test the validity of this approximation on quasi-linear scales using both simulations and observational data  across a broad range of halo masses and redshifts. In particular, we define a separability function based on halo or galaxy cross-correlations to quantify deviations from halo bias separability, and measure it from N-body simulations. We find significant departures from separability on quasi-linear scales (\(\sim 1\text{--}5\,\mathrm{Mpc}\)) at high redshifts (\(z \geq 3\)), leading to a suppression in the scale-dependent halo bias and hence in halo cross-correlations by up to a factor of 2 -- or even higher. In contrast, deviations at low redshifts remain modest. Additionally, using high-redshift (\(z \sim 3.6\)) galaxy samples, we detect deviations from bias separability that closely align with simulation predictions. The breakdown of the separable bias approximation on quasi-linear scales at high redshifts underscore the importance to account for non-separability in models of the galaxy-halo connection in this regime. Furthermore, these results highlight the potential of high-redshift galaxy cross-correlations as a probe for improving the galaxy-halo connection from upcoming large-scale surveys.
\end{abstract}


\section{Introduction}

In hierarchical theory of structure formation,  the gravitationally collapsed dark matter haloes provide potential wells within which galaxies subsequently form and evolve \citep{white_rees_1978}. The spatial clustering of galaxies, as measured by their correlation functions, traces the underlying large-scale structure of the Universe and carries crucial information about the processes governing cosmic structure formation. Clustering measurements, when interpreted within the framework of a theoretical model, serve as a powerful probe to refine models of galaxy-halo connections and galaxy formation \citep{seljak_2000, scoccimarro_2001, bullock+2002, berlind_weinberg_2002, kravtsov+2004, zheng+2005,  lee+2009, zehavi+2011,  harikane2018goldrush, okumura+2021, chaurasiya+2024, yuan+2024, shuntov+2025}. In addition, they impose stringent limits on the growth of structures over cosmic time \citep{yang+2003, conroy+2006, jose+2013, jose+2014, park+2016, bhowmick+2018, jimenez+2019, pei+2024}, and constrain cosmological parameters \citep{tinker+2012_clustering_cosmology, alam+2017, ivanov+2020, abbott+2022, valogiannis+2024, pellejero+2024} and beyond $\Lambda$CDM models \citep{moretti+2023, gsponer+2024, hahn+2024}. 
Current and upcoming large-scale surveys measure galaxy and halo clustering across a broad range of redshifts and galaxy luminosities with unprecedented accuracy, advancing a new era of precision cosmology \citep{lsst_white_paper_212, desi_2016, euclid_2020_cosmology, dalmasso+2024_jwst}.

The correlation functions of dark matter haloes and galaxies are interpreted within the framework of the halo model of large-scale structure \citep{cooray_sheth_2002}. In this model, galaxy clustering arises from two distinct components: the one-halo term, which accounts for clustering within individual haloes, and the two-halo term, which describes clustering between galaxies residing in different haloes. The one-halo term dominates on small scales, where clustering occurs within typical dark matter haloes, whereas the two-halo term is due to the clustering of galaxies in distinct haloes and is important on larger scales. A key quantity for calculating the two-halo term is the \textit{halo bias}, which quantifies the relationship between the spatial distribution of dark matter haloes and the underlying density fluctuations \citep{kaiser_1983_bias, mo_white_1996}. The halo bias is typically calibrated using gravity-only N-body simulations, where physically motivated empirical fitting functions provide accurate approximations across different cosmologies, as theoretical predictions remain challenging (e.g., \citealt{sheth_tormen_1999_bias, Tinker_2005_bias}).

A widely adopted assumption within the halo model is the scale-independent linear halo bias approximation, which applies on large scales \citep{mo_white_1996}. Under this approximation, the halo correlation function is proportional to the matter correlation function, with the proportionality factor given by the square of the scale-independent bias factor \citep{mo_white_1996}. This approximation has been shown to agree well with results from N-body simulations and has proven effective in modelling low redshift galaxy clustering on large scales \citep{zheng+2005, zehavi+2011}. 

On quasi-linear scales, which are a few times larger than the typical sizes of dark matter haloes, the linear bias approximation often underpredicts the clustering strength. This discrepancy, in the simplest approach, can be reduced by incorporating nonlinear matter power spectrum derived from N-body simulations into the halo model (e.g., \citealt{smith+2003}). However, this alone is not sufficient. At low redshifts, where the scale dependence of halo bias on quasi-linear scales is relatively small (typically around ten percent), scale-dependent expressions for halo bias are fine-tuned using N-body simulations and incorporated into the halo model \citep{Tinker_2005_bias, smith+2007, vandenbosch+2013}. At high redshifts, theoretical models and simulations reveal a strong scale dependence of the halo bias, resulting in a significant enhancement of clustering on quasi-linear scales \citep{scannapieco+2002_nlbias, barkana2007_nl_clustering, reed+2007_nlbias}. This effect, which amplifies galaxy clustering by an order of magnitude on quasi-linear scales, has been calibrated using simulations \citep{jose+2016_nlbias} and is essential for accurately interpreting the high redshift galaxy clustering at $z \geq 3$ on quasi-linear scales \citep{jose+2017_lbg_acf, harikane2022goldrush}. 

Another key premise of the linear bias approximation is that the halo bias between two halo populations can be expressed as a separable function of the mass of each population \citep{scoccimarro_2001, smith+2007, smith+2011_nl_clustering, schneider+2012_wdm}. The approximation of bias separability is often adopted in the halo model, as it simplifies the analysis of two-halo clustering \citep{smith+2007}. Incorporating the non-linear matter power spectrum and the scale-dependent halo bias into this framework effectively reproduces observed galaxy clustering across a wide range of redshifts and on large and quasi-linear scales \citep{Zehavi2005, zehavi+2011, vandenbosch+2013,  jose+2017_lbg_acf}.

\citet{mead_verde_2021} introduced a beyond-linear bias framework within the halo model, using a non-separable and scale-dependent halo bias. Independently, earlier theoretical work by \citet{scannapieco+2002_nlbias} had already indicated
that halo bias is not separable at very high redshifts ($z \gtrsim 6$), highlighting the need for more general approaches. A number of subsequent studies have adopted the formalism of \citet{mead_verde_2021} to interpret galaxy clustering data within the halo model \citep{mahony+2022_blb_halomodel, dvornik+2023, pizzati_2024_qso_blb}; however, their primary focus has been on the implications of scale-dependent bias, rather than on explicitly testing bias separability.

In this work, we focus specifically on testing the validity of the separable bias assumption on quasi-linear scales using both cosmological simulations and observational data. We examine this assumption using dark matter simulations that span a wide halo mass range ($5 \times 10^{11}\,M_\odot$--$5 \times 10^{13}\,M_\odot$), which host typical galaxies at both low and high redshifts. Additionally, we perform a clustering analysis of observed bright galaxy samples at $z \sim 3.6$. Our goal is to assess the robustness of the separable bias approximation across redshift and to determine whether its validity can be tested with current high-redshift data.

This paper is organized as follows. In Section~\ref{sec_halo_bias_theory}, we briefly discuss the halo model under the assumption of non-separability of halo bias. Section~\ref{sec_data_simulation} introduces the N-body simulations and observational data used in this study, along with the statistical analysis tools employed. In Section~\ref{sec_results}, we use the halo cross-correlation to assess the validity of the non-separability assumption of the halo bias using simulations, and subsequently apply the same methodology to observational data. The discussion and conclusions are presented in Section~\ref{sec_summary_conclusion}. We adopt the cosmological parameters from \citet{Planck_2018} for all the analysis in this paper. 

\section{Non-Separable halo bias and halo cross-correlations}
\label{sec_halo_bias_theory}
We first introduce the scale-dependent, non-linear halo bias framework for describing the clustering of high-redshift dark matter haloes and galaxies. Since measurements from simulations and observations in this work are made in real space, we present our equations accordingly.

The halo bias $b_{\mathrm{hh}}$ for haloes of masses $M_1$ and $M_2$ separated by a distance $r$ can be defined using the real-space cross-correlation function of haloes, $\xi_{\mathrm{c}}(r, M_1, M_2, z)$, and is given by \citep{cooray_sheth_2002, smith+2007, mead_verde_2021}:
\begin{equation}
\xi_{\rm c}( M_1, M_2, r, z) = b^2_{\rm hh}(M_1, M_2, r, z) \xi^{\rm lin}_{\rm mm}(r, z). \label{eq:cross_corr} 
\end{equation}
Here $\xi^{\rm lin}_{\rm mm}(r, z)$ is the linear matter spatial auto-correlation function. The halo cross-correlation quantifies the excess probability, in N-body simulations, of finding a dark matter halo of one mass $M_1$ at a given distance from a halo of another mass $M_2$, relative to a random distribution, and serves as an important statistic for characterizing the large-scale matter distribution \citep{Peebles1980}. 

By construction, the above definition yields an effective, non-linear halo bias: because $\xi^{\rm lin}_{\rm mm}$ is linear, all non-linear growth of matter clustering on quasi-linear and small scales is absorbed into $b_{\rm hh}(M_1,M_2,r,z)$ in Eq.~\eqref{eq:cross_corr}. One may alternatively define the linear halo bias, $b^{\rm lin}_{\rm hh}$, by replacing $\xi^{\rm lin}_{\rm mm}$ in this equation with the fully non-linear matter correlation $\xi_{\rm mm}$, which removes the non-linear growth of the matter field from the bias definition. As we show later, this choice does not affect the separability statistic introduced in this paper.  

On sufficiently large scales $\xi_{\rm mm} \simeq \xi^{\rm lin}_{\rm mm}$, so both definitions of
bias coincide. In this regime it is common to invoke two further assumptions to simplify calculations. The first is the assumption that the halo bias is a separable function of masses $M_1$ and $M_2$ and is expressed as 
\begin{equation}
    b^2_{\rm hh}(M_1, M_2, r, z) = b(M_1, r, z)  b(M_2, r, z), \label{eq:bias_separable}
\end{equation}
where the scale-dependent halo bias $b(M, r, z)$ of dark matter haloes with mass $M$ is defined via their spatial auto-correlation function as
\begin{equation}
    \xi_{\rm a}(r, M, z) = b^2(M, r, z) \xi^{\rm lin}_{\rm mm}(r, z). 
\end{equation}
The second assumption is the scale independence of halo bias given by
\begin{equation}
    b(M, r, z)  =  b(M, z). \label{eq:bias_scale_independent}
\end{equation}
Both of these approximations hold on sufficiently large scales, where linear perturbation theory provides a highly accurate description of the growth of structures \citep{mo_white_1996, cooray_sheth_2002}. Under these conditions, Eq.~\eqref{eq:cross_corr} simplifies to
\begin{equation}
\xi_{\rm c}( M_1, M_2, r, z) \simeq b(M_1, z) b(M_2, z) \xi^{\rm lin}_{\rm mm}(r, z), \label{eq:lin_bias_approx}
\end{equation}
where $b(M,z)$ is the scale-independent linear halo bias.

Several prescriptions for the scale-independent halo bias have been proposed in the literature, beginning with analytic models (e.g.~\citealt{mo_white_1996}), and subsequently refined into accurate fitting functions calibrated using N-body simulations. A commonly used parametrization is that of \citet{Tinker+2010_bias}, who proposed the fitting formula
\begin{equation}
b(M,z) \equiv  b\left[\nu(M,z) \right] = 1 - A\,\frac{\nu^{a}}{\nu^{a} + \delta_{c}^{a}}
         + B\,\nu^{b} + C\,\nu^{c},
\label{tinker_bias_equ}
\end{equation}
where $A$, $B$, $C$, $a$, $b$, and $c$ are parameters determined empirically from simulations.

In these models, $b(M,z)$ is expressed as a function of the peak height $\nu(M,z) = \delta_c / \sigma(M,z)$, which quantifies the rarity of dark matter haloes \citep{kaiser_1983_bias, mo_white_1996, cooray_sheth_2002}. Here $\delta_c = 1.686$ is the critical overdensity for spherical collapse, and $\sigma(M,z)$ is the root-mean-square (rms) linear density fluctuation on the mass scale $M$, defined as
\begin{equation}
\sigma^2(M, z) = \int_0^\infty \frac{dk}{2\pi^2}\, k^2
P^{\rm lin}_{\rm mm}(k, z)\, W^2(k, R),
\label{eqn:sigma}
\end{equation}
where $R$ is the comoving radius enclosing mass $M$, $W(k,R)$ is the Fourier transform of the spherical top-hat window function, and $P^{\rm lin}_{\rm mm}(k,z)$ is the linear matter power spectrum, i.e.\ the Fourier transform of $\xi^{\rm lin}_{\rm mm}$ \citep{Peebles1980}.

Among these two assumptions, the scale dependence of halo bias has been more extensively studied using analytic models and N-body simulations. Several alternative models have been developed to better describe how halo bias varies with scale, particularly on quasi-linear scales ranging from approximately 0.5 Mpc to a few Mpc. We therefore briefly discuss this aspect before returning to the validity of the assumption of separability of bias. 

Theoretical and observational efforts have shown that halo bias exhibits a pronounced scale dependence on quasi-linear scales—those larger than typical halo sizes \citep{scannapieco+2002_nlbias, iliev+2003_nlclustering, Tinker_2005_bias, vandenbosch+2007, barkana2007_nl_clustering, reed+2007_nlbias, jose+2016_nlbias, jose+2017_lbg_acf, harikane2022goldrush}. 
In particular, both the linear bias and the non-linear bias (as defined in Section~\ref{sec_halo_bias_theory}) exhibit scale dependence on these quasi-linear scales, with the latter definition showing a stronger scale dependence because it absorbs the non-linear growth of the matter field \citep{vandenbosch+2007,jose+2016_nlbias}.
At low redshifts, this effect leads to a modest increase in the correlation function, typically by a few tens of percent \citep{Tinker_2005_bias}. On the other hand, at high redshifts, it can significantly amplify halo clustering \citep{jose+2016_nlbias}. For rare haloes with masses exceeding \(10^{12} \rm M_\odot\) at $z \geq 3$, the bias becomes strongly scale-dependent, leading to a correlation function that can be up to an order of magnitude higher on quasi-linear scales than predicted by the linear bias model. 

The separability of halo bias, as expressed in Eq.~\eqref{eq:bias_separable}, has been more widely adopted than scale-independence of the bias in analytic models of galaxy clustering \citep{smith+2007, smith+2011_nl_clustering, lap_2021_stochBias}. While computationally simple and often sufficiently accurate, this simplification has recently been revisited. In particular, \citet{mead_verde_2021} proposed an alternative beyond linear bias framework where the halo bias is scale-dependent and no longer a separable function of $M_1$ and $M_2$. In this case, the halo power
spectrum is defined as, 
\begin{equation}
\begin{aligned}
   P(M_1, M_2,  k, z) &= b(M_1, z) b(M_2, z) P^{\rm lin}_{\rm mm}(k,z) \\ 
   &\quad \times  \left[ 1 + \beta_{\text{NL}}(M_1, M_2, k,z) \right], \label{eq_mead_bias}   
\end{aligned}  
\end{equation}
The term $\beta_{\text{NL}}(M_1, M_2, k,z)$ is a non-separable function of $M_1$ and $M_2$ that also incorporates the scale dependence of halo bias. On large scales, where the linear bias approximation remains valid, $1+\beta_{\text{NL}}$ approaches unity. 

At $z \sim 0$, for haloes with masses in the range of $10^{11} - 10^{13} M_\odot$, $\beta_{\text{NL}}$ enhances halo cross-correlations by 10-20\% at scales $k \sim 0.5$ h/Mpc with the enhancement vanishing on larger scales ($k < 0.04$ h/Mpc). Following \citet{mead_verde_2021}, a number of studies have incorporated this beyond-linear bias into the halo model to understand the observed clustering of galaxies at low redshifts \citep{mahony+2022_blb_halomodel, dvornik+2023} and analyze the quasar correlation functions at $z \sim 6$  \citep{pizzati_2024_qso_blb}.  

In Eq.~\eqref{eq_mead_bias}, \citet{mead_verde_2021} treat the scale dependence and mass non-separability of the halo bias jointly, rather than examining the individual contributions of these two effects. To assess whether the assumption of bias separability remains valid on quasi-linear scales, we factor out the scale dependence of the non-linear bias measured for haloes of a given mass and define a real-space halo bias separability function $s(M_1, M_2, r, z) = 1 + \alpha(M_1, M_2, r, z)$ as: 
\begin{equation}
\begin{aligned}
\xi_{\rm c}(M_1, M_2, r, z) &= b(M_1, r, z) \, b(M_2, r, z)  \xi_{\rm mm}^{\rm lin}(r) \,  \\
            & \quad \times \left[ 1 + \alpha(M_1, M_2,  r, z) \right].
\end{aligned}
\end{equation}

Note that while \( s \) retains a dependence on scale, the explicit scale dependence associated with the halo bias at a given mass, \( b(M, r, z) \), has been separated out by this definition. The function $s(M_1, M_2, r, z)$ therefore captures the additional scale dependence present in the halo cross-correlation of unequal-mass pairs relative to that inferred from the corresponding auto-correlations at each mass. The separability function $s(M_1, M_2, r)$ can be estimated from cosmological N-body simulations as 
\begin{equation}
s(M_1, M_2, r,z) = \frac{\xi_{\rm c}(M_1, M_2,  r,z)}{\sqrt{\xi_{\rm a}(M_1, r,z) \, \xi_{\rm a}(M_2, r,z)}}.
\label{eq_alpha}
\end{equation}
By construction, $s$ involves only ratios of measured non-linear halo correlation functions and therefore does not depend on whether one chooses to define an effective bias using $\xi^{\rm lin}_{\rm mm}$ or the fully non-linear $\xi_{\rm mm}$. As a result, $s$ probes the separability of halo clustering independently of the bias normalisation convention. 

On sufficiently large scales, where the linear bias approximation is valid, the function $s(M_1, M_2, r, z)$ is expected to approach unity, similar to the behaviour of $1 + \beta_{\text{NL}}$, indicating that the halo bias becomes effectively separable in $M_1$ and $M_2$. By construction, $s(M_1, M_2, r, z)$ also tends to unity when the two halo masses become equal ($M_1 \to M_2$), and this holds across all spatial scales. In contrast, $1 + \beta_{\text{NL}}$ can deviate from unity even as $M_1 \to M_2$, particularly on quasi-linear scales, because it also incorporates the scale dependence of the bias itself. Thus, $s$ is primarily sensitive to the non-separability of  scale-dependent halo bias across different halo masses. Therefore, the function $s(M_1, M_2, r, z)$ provides a direct diagnostic of whether the scale dependence of the non-linear halo bias factorizes with respect to halo mass.

Furthermore, the estimation of $\beta_{\text{NL}}$ from observations requires knowledge of the linear matter correlation function, which in turn depends on the assumption of a specific set of cosmological parameters. In contrast, the function $s$ can, in principle, be measured directly from observations, as it requires only halo–halo (or galaxy–galaxy) cross- and auto-correlation functions for its computation. This make $s$ less reliant on cosmological modelling assumptions, enabling for a more straightforward comparison between observations and simulations.

\subsection{How rare are high-z galaxy-hosting haloes?}

\begin{figure}
     \centering
           \includegraphics[width=1.0\linewidth]{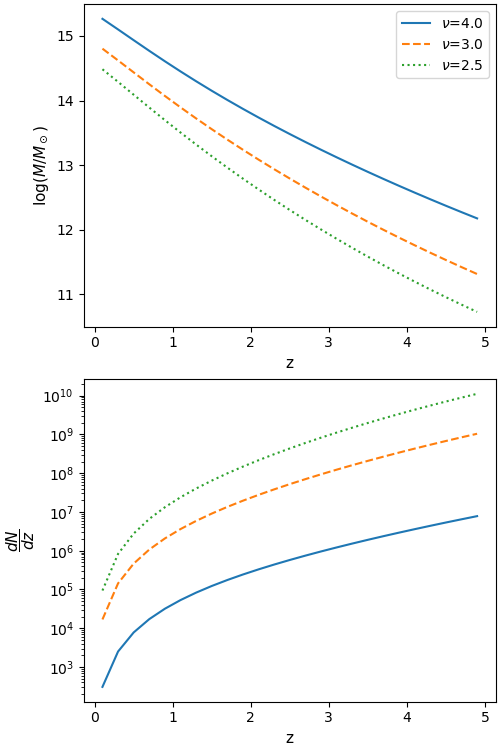}
           \caption{Top panel: Halo masses corresponding to various values of $\nu$ as a function of redshift. Bottom panel: The total number of dark matter haloes in unit logarithmic mass interval in the sky, per unit redshift interval, with a given $\nu$, as a function of redshift.
} \label{fig_halo_rarity}
\end{figure} 

It is important to note that the high-redshift galaxy populations used in this study are statistically distinct from their low-redshift counterparts in terms of their host halo properties. The galaxies analysed here have halo masses ranging from approximately $5 \times 10^{11}$ to a few times $10^{13} \,\rm M_\odot$ \citep{harikane2022goldrush}. At low redshift $(z \sim 0)$, emission-line galaxies and luminous red galaxies (LRGs) from SDSS and DESI, which are typically not associated with clusters or groups, have comparable halo masses \citep{masaki_2014_lrg, vanUitert_2015, gao_2023_lrg_elg_DESI, berti+2023_sham}. 

Although the halo masses of high-redshift galaxies are slightly smaller than their low-redshift counterparts, high-redshift haloes are statistically much rarer, as they form from high-$\sigma$ fluctuations. Using Eq.~\eqref{eqn:sigma} and $\nu = \delta_c/\sigma$, we obtain the masses of haloes collapsing from $2.5$, $3$, and $4\sigma$ fluctuations as a function of redshift, which we show in the top panel of Fig.~\ref{fig_halo_rarity}. It is evident that, for a given $\nu$, the halo mass is lower at high $z$, which means that haloes of a given mass are much rarer at high redshifts. For example, $10^{13} \, \rm M_\odot$ haloes at $z \sim 4$, which collapse from $4\sigma$ fluctuations, are as rare as superclusters at  $z \sim 0$.  

In the lower panel of Fig.~\ref{fig_halo_rarity}, we also show the total number of dark matter haloes per unit logarithmic mass and redshift in the sky, corresponding to a given $\nu$, obtained from the \citet{Tinker_2005_bias} halo mass function. Clearly, for a given $\nu$, this number increases with redshift. For example, at $z \sim 4$, there are about a million haloes with a mass of $10^{13} \, \rm M_\odot$ per unit logarithmic mass interval and unit redshift, compared to only a few hundred objects with similar $\nu$ ($M \sim 5 \times 10^{15} \,\rm M_\odot$) at $z \sim 0$. These larger samples of rare, high-redshift haloes enable detailed statistical analyses of their properties. As tracers of the extreme tail of the halo mass distribution, their statistical behaviour is expected to differ significantly from that of their low-redshift counterparts.

\section{The data and Measurements}
\label{sec_data_simulation}

In this section, we provide an overview of the N-body simulations and observational data used in our study to probe the impact of non-separable bias on quasi-linear scale clustering of dark matter haloes and galaxies. Additionally, we describe the statistical estimators employed for measuring halo and galaxy correlation functions, along with the methods used to quantify uncertainties.

\subsection{The Abacus N-body Simulations}

We use the AbacusSummit suite of N-body simulations, a large volume, high-accuracy and high-resolution dataset designed to support large-scale structure studies in the era of next-generation surveys~\citep{abacussummit_maksimova2021}. It consists of over 150 simulations covering 97  cosmological models. These simulations were produced using Abacus, a GPU-optimized N-body code that employs the static multipole mesh method to efficiently compute gravitational interactions \citep{abacus_garrison2021,hadzhiyska2022compaso}. They enable precise measurements of halo auto- and cross-correlations, particularly for very massive dark matter haloes at high redshifts, which are required in this study.

To compute halo correlations, we use simulations based on the Planck 2018 baseline $\Lambda$CDM cosmology~\citep{aghanim+2018}. In particular two simulation volumes are used: (i) huge simulation, with $8640^{3}$ particles in 7.5 $h^{-1}$ Gpc box, with a particle mass resolution of $5 \times 10^{10} h^{-1}\rm M_{\odot}$, and (ii) base simulation, with $6912^{3}$ particles in 2$h^{-1}$Gpc box, with a particle mass resolution of $2 \times 10^{9} h^{-1}\rm M_{\odot}$\citep{abacussummit_maksimova2021}. Dark matter haloes in these simulations are identified using the CompaSO algorithm ~\citep{hadzhiyska2022compaso}, which assigns halo masses based on spherical overdensity criteria. We adopt $M_{200}$, the mass enclosed within a region of 200 times the critical density of the universe, as our halo mass definition. 

\subsection{The HSC-SSP observations}
\label{sec_hsc_ssp_data}

\begin{figure}
    \centering
    \includegraphics[width=0.9\linewidth]{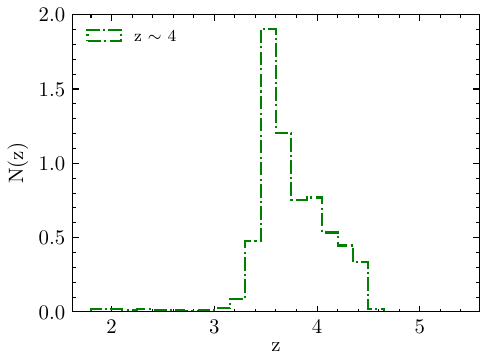}
    \caption{Normalized redshift distribution of g-dropouts with $20<\rm m_{\rm UV}<24.5$ from HSC-SSP wide survey. }
    \label{fig_Nz}
\end{figure}

Our analysis employs galaxy catalogues from the third public data release (PDR3) of HSC-SSP, as presented in \citet{aihara+2022_pdr3_subaru} (see \citealt{aihara+2018_pdr1_subaru, aihara2019second} for earlier data releases). The HSC-SSP survey spans three layers —wide ($\sim$ 670 deg$^2$), deep ($\sim$ 28 deg$^2$), and ultra-deep ($\sim$ 3 deg$^2$). In this work, we utilize wide layer galaxies from the Spring and Autumn fields covering approximately 600 deg$^2$. Photometric data from the five HSC broad-band filters ($g$, $r$, $i$, $z$, $y$) \citep{kawanomoto2018hyper} are processed through the \textsc{hscPipe} reduction pipeline \citep{Bosch+2018_hscpipe} to generate the galaxy catalogues. These catalogues are used to measure the auto- and cross-correlations of magnitude-selected Lyman Break Galaxy (LBG) g-dropout samples which are identified via the Lyman break technique  which traces the redshifted Lyman-limit break (at rest-frame wavelength of 912~\AA) in the galaxy spectral energy distributions \citep{steidel1996spectroscopic, giavalisco2002lyman}.

As in \citet{harikane2022goldrush}, we select galaxies with signal-to-noise ratios greater than $5$ within $1.5''$ diameter apertures and apply the masks, threshold values, and flags described in Section 2 and Table 2 of \citet{harikane2022goldrush} to exclude galaxies compromised by pixel issues, cosmic rays, and bright source haloes \citep{coupon2018bright}. The color selection for dropout galaxies is then applied to the convolved fluxes within $2''$ diameter apertures after aperture correction, specifically for g-dropouts as outlined in \citet{ono2018great}, \citet{toshikawa2018goldrush}, and \citet{harikane2018goldrush}.

    \begin{equation}
    ( g - r > 1.0 ) \Lambda  ( r - i < 1.0 ) \Lambda  ( g - r > 1.5 ( r - i ) + 0.8 ).  
    \end{equation}

Once the galaxies are color-selected, the fixed $2''$ diameter aperture magnitude after aperture correction is used as the total magnitude in all filters \citet{harikane2022goldrush}. The UV magnitude ($\text m_{\text {UV}}$) for the g-dropouts is the $i$-band magnitude, whose central wavelength is the nearest to the rest frame wavelength of $\lambda \approx 1500$ \AA.

\subsection{Removal of low-z interlopers}
Low-z interlopers satisfying the above color-selection criteria contaminate the HSC-SSP. In order to remove these, we utilize the photometric redshifts (photo-z) of the galaxies as given by PDR3, derived using the empirical photometric redshift fitting code DEmP \citep{hsieh2014estimating, tanaka2018photometric}. The DEmP utilizes template data of galaxies with spectroscopic redshifts from the literature to obtain a polynomial fit to the probability distribution function (PDF) of photo-z of galaxies. This fit is then used to derive the best-fit value, which we adopt in this work.

As in \citet{toshikawa+2024}, the low redshift interlopers are also removed from the g-dropouts, by selecting g-dropout galaxies with the upper bound on the 95 percent confidence interval of the photo-z greater than $2.8$. We show in Fig.~\ref{fig_Nz}, the distribution of the best-fit photo-z, $N(z)$, of the g-dropouts with $20<\rm m_{\rm UV}<24.5$. This constitutes our final sample, which will be used for clustering analysis. The median redshift of the final sample is $z_m \sim 3.6$.


\subsection{The 2-point auto and cross-correlations}
\label{sec_acf_ccf}
In this section, we discuss how to compute the angular or spatial two-point auto- and cross-correlations of galaxy or halo samples discussed in previous sections. The angular two-point auto-correlation is estimated using the well-known Landy-Szalay estimator \cite{landy1993bias} as  
\begin{equation}
            \omega_{\rm a} (\theta)= \frac{\langle DD(\theta)\rangle-\langle 2DR(\theta)\rangle+\langle RR(\theta)\rangle}{\langle RR(\theta)\rangle}\,  
            \label{eq_wtheta}
\end{equation}
where $\langle DD(\theta) \rangle$, $\langle DR(\theta)\rangle$, and $\langle RR(\theta)\rangle$ are the numbers of galaxy-galaxy, galaxy-random, and random-random pairs at a pair separation angle $\theta$, normalized by the total number of pairs. A random catalogue that accounts for masked regions and edge effects within the survey area is essential for computing the number of random pairs. In this work, we have used the random catalogues supplied by \cite{aihara+2022_pdr3_subaru} (see also \citealt{coupon2018bright}), with a surface density of 100 points per arcmin$^2$. 

To study the clustering between two distinct galaxy samples (sample 1 and 2), we compute their two-point angular cross-correlation function using 
\begin{equation}
            \omega_{\rm c}(\theta)= \frac{\langle D_1D_2(\theta)\rangle-\langle D_1R(\theta)\rangle-\langle D_2R(\theta)\rangle
            +\langle RR(\theta)\rangle}{\langle RR(\theta)\rangle}\, , 
            \label{eq_wtheta_cross}
\end{equation}
where $\langle D_1D_2(\theta) \rangle$ is the normalized number of galaxy-galaxy pairs between galaxy sample 1 and 2  at a pair separation angle $\theta$. Similarly, $\langle D_1R(\theta)\rangle$ is the number of galaxy-random pairs between galaxy sample 1 and random catalogue. In this work, we obtain both samples by selecting galaxies from two non-overlapping magnitude bins. Although these bins are separated in apparent magnitude, the broad redshift range of the sample, together with the intrinsic scatter between UV luminosity and halo mass, implies that some overlap in the underlying galaxy masses is possible in principle.

The spatial auto-correlation and cross-correlation functions of dark matter haloes are computed similarly using Eqs.~\eqref{eq_wtheta} and \eqref{eq_wtheta_cross}, but the pair count is measured as a function of real-space separation $r$ of haloes instead of their angular separation $\theta$. Thus, in this case, $\langle DD(r) \rangle$, $\langle DR(r) \rangle$, and $\langle RR(r) \rangle$ correspond to the halo-halo, halo-random, and random-random pair counts, respectively. 
\subsubsection{Statistical Uncertainties}\label{sec_Statistical Uncertainties}
The statistical errors of the angular correlation function are estimated using the Jackknife resampling method \citep{norberg2009statistical} with Jackknife regions of size $\sim 5$ square degrees. The Jackknife covariance matrix is then 
\begin{equation}
          C_{ij} =\frac{ N-1}{N}\sum_{k=1}^{N} (\omega^k (\theta_i)-\bar\omega (\theta_i))(\omega^k (\theta_j)-\bar\omega (\theta_j))
         \end{equation}
where $\omega^k (\theta_i)$ is the angular correlation function at $\theta$=$\theta_i$ from the $k^{th}$ Jackknife region and 
$\bar\omega$ is the average correlation function of a total of $N$ Jackknife regions \citep{scranton2002analysis, Zehavi2005}.

 \begin{figure*}
     \centering
           \includegraphics[width=0.95\linewidth]{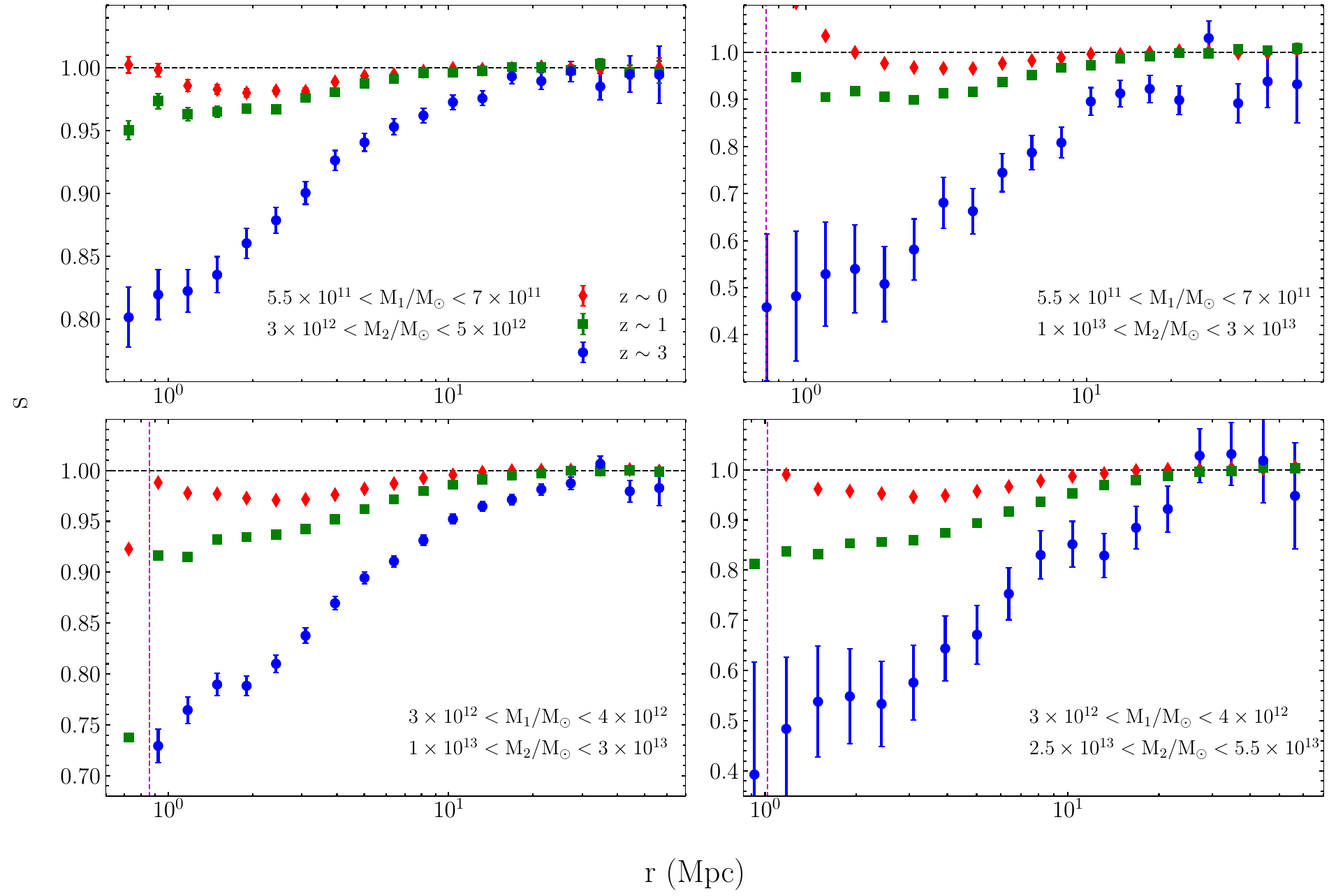}
           \caption{ The halo bias separability function as a function of the halo separation at the redshifts 0 (red diamonds), 1 (green squares) and 3 (blue circles), for halo samples in different mass bins, as indicated in the legend. The measurements in upper and lower panels are respectively from the base and the huge simulations.}
    \centering
           \label{fig_s_simulation}
 \end{figure*} 

\begin{table*}
\begin{tabular}{|l|l|l|l|}
\hline
\begin{tabular}[c]{@{}l@{}} Magnitude \\ ($m_{\rm UV}$) bins \end{tabular} & 20 - 23.9 & 24.2 - 24.5 & 20 - 23.75  \\ \hline
\begin{tabular}[c]{@{}l@{}}Number of \\ galaxies\end{tabular} & 19306 & 99943 & 11034 \\ \hline
\begin{tabular}[c]{@{}l@{}}Effective bias\\ $({\text b_{\text{eff}}})$\end{tabular} & 9.46±0.69 & 5.06±0.19 & 11.8±1.2 \\ \hline
\begin{tabular}[c]{@{}l@{}}Effective mass\\ $({\text M_{\text {eff}}/\text M_\odot})$\end{tabular} & $9.86\substack{+2.1 \\ -1.9}\times 10^{12}$ & $1.39\substack{+0.19 \\ -0.17}\times 10^{12}$& $1.8\substack{+0.53 \\ -0.46}\times 10^{13}$  \\ \hline
\end{tabular}
\caption{Number of g-drop galaxies in different magnitude subsamples, and the corresponding effective bias and halo mass.}\label{table1}
\end{table*}

\section{Result and Discussion}
\label{sec_results}

 \subsection{Non-separable bias from Simulations} 
\label{sec_s_simul}
In this section, we present the halo bias separability function, measured from N-body simulations at different redshifts. To measure $s (M_1, M_2, r,z)$, we require two halo samples with different masses. Therefore, we select haloes from the simulation in a pair of distinct mass bins; the first bin, denoted by $[M_1]$, is defined as $M_1^\text{start} \leq M_1 < M_1^\text{end}$. The average mass of the bin $[M_1]$ is then taken to be the geometric mean $M_1^\text{av} = \sqrt{ M_1^\text{start}  M_1^\text{end}}$. The corresponding average virial radius is given by, 
\begin{equation}
    r_1^\text{av} = \left(\dfrac{3M_1 ^\text{av}}{4 \pi \rho_\text{cric} \Delta}\right)^{1/3},
    \label{eq_rvir}
\end{equation}
where $\rho_\text{cric}$ is the critical density of the Universe and $\Delta=200$. Similarly, for the second sample $[M_2]$, we have $M_2^\text{start} \leq M_2 < M_2^\text{end}$. The halo masses in the two samples are chosen to be non-overlapping.  
We also define the effective mass scale of the bin pair as $M^{\rm av} \equiv \sqrt{M_1^{\rm av} M_2^{\rm av}}$, 
and the logarithmic mass separation between the bins as $\Delta \log M^{\rm av} \equiv \log(M_2^{\rm av}/M_1^{\rm av})$, which together characterize the location and separation of the mass bins.

\begin{figure*}
    \centering
    \includegraphics[width=0.95\linewidth]{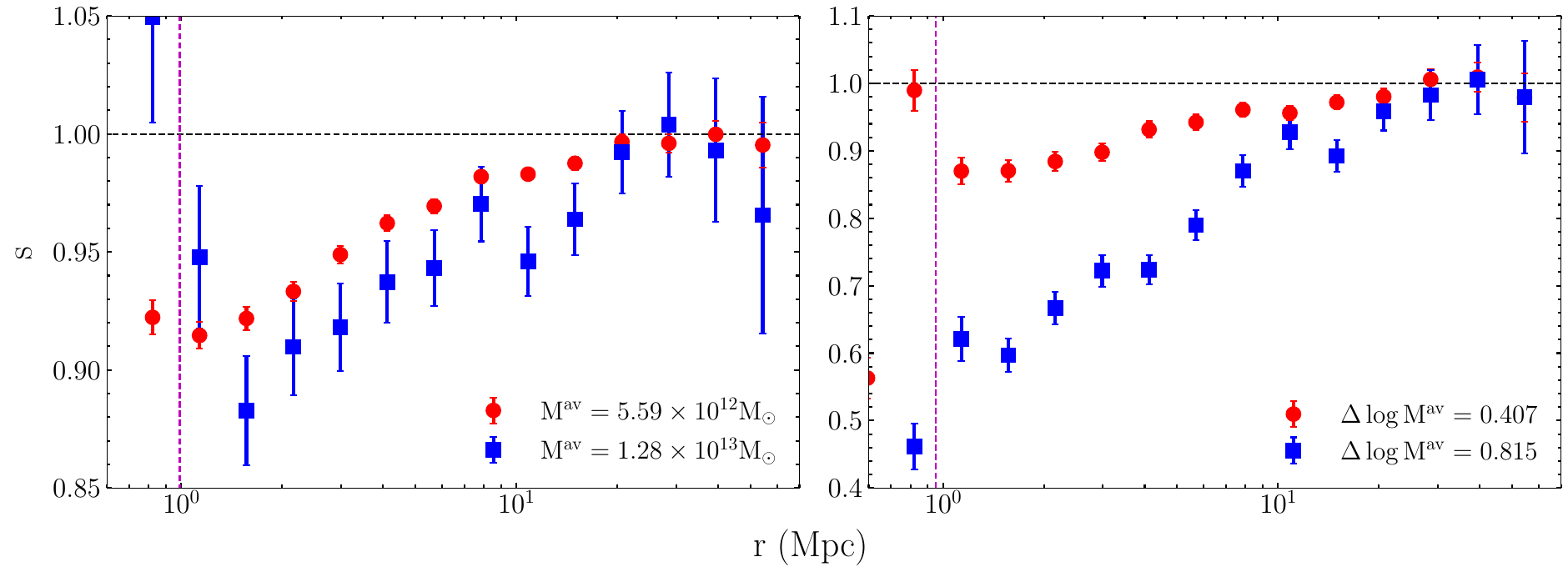}
    \caption{
The dependence of  $s(r)$  on effective mass of bin pairs, $M^{\rm av}$, and logarithmic mass separation between bins, $\Delta \log M^{\rm av}$. \emph{Left panel:} $s(r)$ measured for two halo pairs with the same $\Delta \log M^{\rm av} \simeq 0.36$ but different $M^{\rm av}$.  Red circles correspond to mass bins $3.0\times10^{12} < M_1/M_\odot < 4.55\times10^{12}$ and $6.89\times10^{12} < M_2/M_\odot < 1.04\times10^{13}$, while blue squares correspond to $6.89\times10^{12} < M_1/M_\odot < 1.04\times10^{13}$ and $1.58\times10^{13} < M_2/M_\odot < 2.40\times10^{13}$. The $M^{\rm av}$ of bin pairs are $5.59\times10^{12}\,M_\odot$ (red squares) and $1.28\times10^{13}\,M_\odot$ (blue squares). 
\emph{Right panel:} $s(r)$ measured for two halo pairs with the same $M^{\rm av} \simeq 10^{13}\,M_\odot$ but different $\Delta \log M^{\rm av}$. Red circles corresponds to $5.21\times10^{12} < M_1/M_\odot < 7.50\times10^{12}$ and
$1.33\times10^{13} < M_2/M_\odot < 1.92\times10^{13}$, while blue squares correspond to $3.26\times10^{12} < M_1/M_\odot < 4.69\times10^{12}$ and $2.13\times10^{13} < M_2/M_\odot < 3.07\times10^{13}$.  Here $\Delta \log M^{\rm av} =0.407$ for the red circles and  $\Delta \log M^{\rm av} = 0.815$ for the blue squares.
}
\label{fig_s_mass_trends}
\end{figure*}

We then estimate the auto-correlation function, $\xi_\text{a}([M], r,z)$, of dark matter haloes within each mass bin, as well as the cross-correlation, $\xi_\text{c}([M_1], [M_2], r,z)$, between haloes in the two bins. These correlation functions are used to compute the separability function $s$ (as defined in Eq.~\eqref{eq_alpha}). A deviation of $s$ from unity quantifies the extent to which halo bias is non-separable across the selected mass bins.

Fig.~\ref{fig_s_simulation} shows the separability function $s$ measured for different pairs of halo mass bins at  $z \sim 0$, 1, and 3, as a function of separation $r$. Each panel presents the evolution of $s$ with redshift for a fixed pair of mass bins, which are specified in the legend.  For example, in the first panel, $s$ is measured using halo mass bins defined by $5.5 \times 10^{11} < M_1/M_\odot < 7 \times 10^{11}$ and $3 \times 10^{12} < M_2/M_\odot < 5 \times 10^{12}$.  We use relatively wider mass bins to ensure that each sample contains a sufficient number of haloes for a statistically robust measurement of $s$. The vertical lines in each panel indicate a conservative upper limit on the halo exclusion scale, $r_1^{av} + r_2^{av}$. Below this scale, halo exclusion effects can be important; therefore, we do not analyze the behavior of $s$ in that regime.

We first note that for all pairs of mass bins, $s$ departs from unity on quasi-linear scales (approximately $1\text{--}5$ Mpc), demonstrating that halo bias is a non-separable function of $M_1$ and $M_2$ in this regime. On most of these scales, and at all redshifts, $s$ falls below unity, which implies an reduced probability of finding haloes of masses $M_1$ and $M_2$ at a separation $r$ relative to the expectation from Eqs.~\eqref{eq:cross_corr} and \eqref{eq:bias_separable}. However, at $z \sim 0$ and on scales around $1 \mathrm{Mpc}$, $s$ slightly exceeds one, suggesting a small enhancement in the likelihood of such halo pairs being found at that distance. 

The extent to which the separability factor differs from unity varies significantly with redshift. At low redshifts $( z \sim 0 )$, this difference remains relatively modest---at most $ \sim 10\% $ for the mass ranges considered here. At an intermediate redshift of $z \sim 1$, the maximum deviation increases to $\sim 20\%$, indicating a stronger departure from separability. In contrast, at higher redshifts, the shift from unity becomes substantially more pronounced.  At $z \sim 3$, $s$ deviates considerably from unity, reaching values as low as $\sim 0.45$ (see the right panels of Fig.~\ref{fig_s_simulation}). This strong deviation clearly indicates that the assumption of separable halo bias breaks down at high redshifts ($z \gtrsim 3$) on quasi-linear scales, resulting in a reduction of the cross-correlation function by the same factor.  Since these scales are significantly larger than our conservative estimate of the halo-exclusion scale, this effect cannot be attributed to halo exclusion.

The scale up to which $s$ departs from unity also exhibits a redshift dependence. At $z \sim 0$, non-separable bias remains significant up to scales of approximately $5-10 \, \mathrm{Mpc}$, beyond which $s$ approaches unity, indicating that the halo bias becomes effectively separable. In contrast, for halo samples of similar mass at $z \sim 3$, the deviation from unity persists even on scales of about $20 \, \mathrm{Mpc}$. This demonstrates that the scale over which the bias is non-separable increases with redshift for given $M_1$ and  $M_2$.

The separability function $s$ becomes more pronounced for higher-mass haloes and for halo pairs drawn from more widely separated mass bins at a fixed redshift. To illustrate these, Fig.~\ref{fig_s_mass_trends} compares $s(r)$ for different combinations of effective halo mass of bin pairs, $M^{\rm av}$,  and logarithmic mass separation between bins,  $\Delta \log M^{\rm av}$. In the left panel of Fig.~\ref{fig_s_mass_trends}, we show $s(r)$ for two pairs of halo mass bins that share the same $\Delta \log M^{\rm av}$ but differ in their $M^{\rm av}$. At fixed $\Delta \log M^{\rm av}$, the deviation of $s$ from unity is more pronounced for larger $M^{\rm av}$, indicating that bias non-separability is enhanced for more massive haloes.

In the right panel, we instead fix the effective mass scale of bin pairs and compare halo pairs with different $\Delta \log M^{\rm av}$. In this case, the deviation of $s$ from unity increases systematically with increasing $\Delta \log M^{\rm av}$, demonstrating that non-separability increases as the mass separation between the two halo populations grows.

The deviation of the separability function $s$ from unity quantifies the extent to which the scale dependence of halo clustering fails to factorize with respect to halo mass. Since $s<1$ on quasi-linear scales, the scale dependence of halo cross-correlations becomes weaker as the mass separation between halo populations increases. Consequently, the scale-dependent bias inferred from halo auto-correlations does not provide an accurate description of the cross-correlation of unequal-mass haloes. This effect becomes increasingly pronounced at higher redshifts, evolving from a modest $5$--$10\%$ deviation at low redshift to a suppression by a factor of $\sim 2$ by $z \sim 3$.

As discussed in Section~\ref{sec_halo_bias_theory}, massive haloes at high redshift correspond to rare $3$--$4\sigma$ peaks of the initial density field, whose joint clustering properties are expected to differ  from those of more common haloes at lower redshifts. This behaviour is qualitatively consistent with theoretical expectations from excursion-set--based analyses of the joint statistics of haloes of different masses by \citet{scannapieco+2002_nlbias}. In their framework, the bivariate mass fraction that describes the fraction of collapsed haloes of masses $M_1$ and $M_2$ at a fixed separation is not, in general, a separable function of the corresponding single-point mass fractions. Moreover, this non-separability becomes particularly important at high redshift and for massive haloes. However, these studies did not explicitly examine whether the resulting scale-dependent bias exhibits a suppression relative to the separable bias expectation, as quantified here by $s<1$. A detailed quantitative comparison between the separability function measured in this work and predictions from excursion-set theory is therefore left for future investigation.

\subsection{Non-separable bias from Galaxy Cross-Correlation Functions}

 \begin{figure*}
     \centering
           \includegraphics[width=0.95\linewidth]{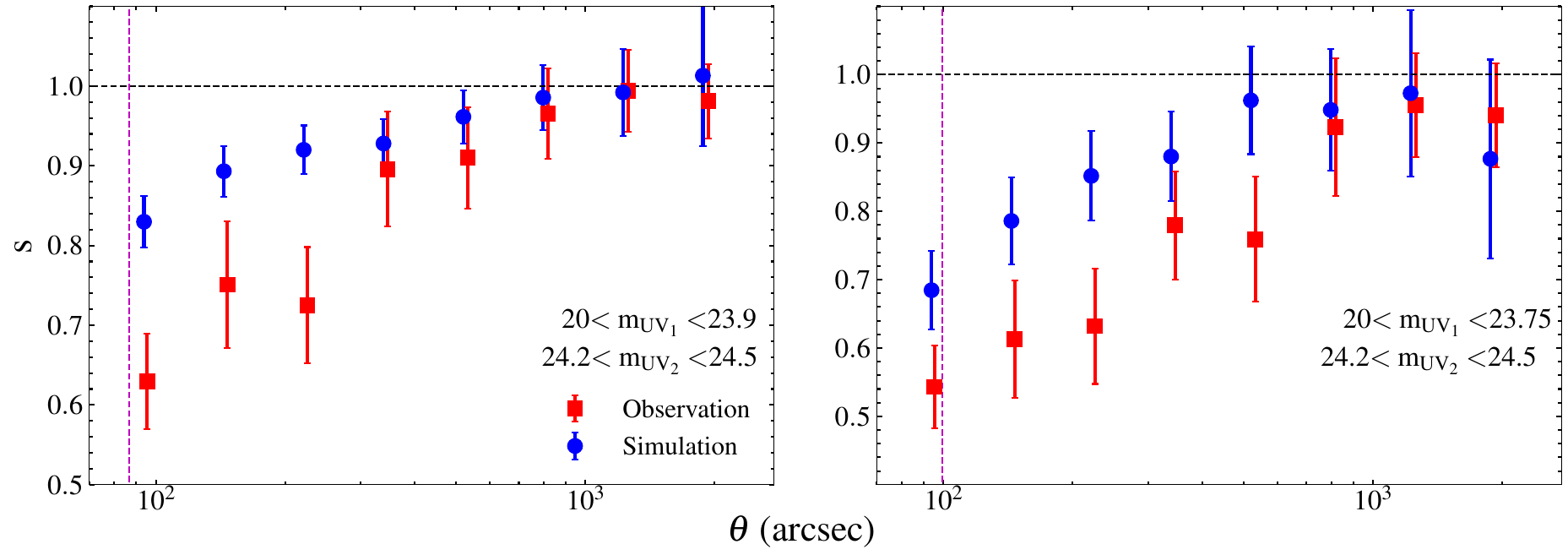}
           \caption{The bias separability function derived using HSC-SSP galaxy samples at $z \sim 3.6$ (red squares)  and Abacus Simulation halo catalogues at $z \sim 3$ (blue circles) using base simulation. 
           \label{fig_cf_ratio}}
 \end{figure*}

 The strong deviation of the separability function from unity at high redshift, found in the previous section, presents a unique opportunity to measure the same from observations. In this section, we measure the separability function using galaxy samples from the HSC-SSP survey at $z \sim 3.6$ as a function of angular scale $\theta$.

These high redshift galaxy populations from the HSC-SSP survey are typically associated with halo masses exceeding $10^{12} {\rm M}_\odot$ \citep{harikane2022goldrush}, placing them in a regime where non-separability is expected to be very prominent. Moreover, the large number of galaxies from the 600 deg$^2$ wide field makes this dataset particularly well-suited for measuring $s$ with high statistical significance.

In order to measure $s$, we divide the galaxy sample into two UV magnitude bins, denoted by $[\rm m_\text{uv1}]$ and $[\rm m_\text{uv2}]$. Theoretical modelling of LBG luminosity functions and clustering indicates that rest-frame UV luminosity correlates with halo mass, with an intrinsic scatter in UV luminosity at fixed halo mass. Combined with the fact that the magnitude-binned samples spans a broad redshift range (see Fig.~\ref{fig_Nz}), this implies some overlap in the underlying halo masses of galaxies selected in different UV-luminosity bins. Nevertheless, existing studies suggest that, in the halo mass range relevant for our analysis, the scatter in LBG UV luminosity is modest (e.g. \citealt{giavalisco+2001}; see also \citealt{lee+2009}), such that luminosity-based binning provides a reasonable proxy for selection by halo mass.
We then measure the angular auto-correlation of galaxies within each magnitude bin and the angular cross-correlation between galaxies in these two bins. Using these measurements, we calculate the separability function as 
\begin{equation}
 s ([\rm m_\text{uv1}], [\rm m_\text{uv2}], \theta, z) = \frac{\omega_{\rm c}([\rm m_\text{uv1}], [\rm m_\text{uv2}], \theta, z)}{\sqrt{\omega_{\rm a}([\rm m_\text{uv1}], \theta, z) \, \omega_{\rm a}([\rm m_\text{uv2}], \theta, z)}}. 
    \label{eq:s_def}
\end{equation}
In Fig.~\ref{fig_cf_ratio}, we present the measured separability function $s$ (red squares) for two cases, each constructed from a pair of non-overlapping UV magnitude bins. In the first case, shown in the left panel, the bins are defined as $20 < \rm m_\text{uv1} \leq 23.9$ and $24.2 < \rm m_\text{uv2} \leq 24.5$. In the second case, shown in the right panel, the magnitude gap between bins is slightly wider, with the bins given by $20 <\rm m_\text{uv1} \leq 23.75$ and $24.2 <\rm m_\text{uv2} \leq 24.5$.  Thus the samples used in the two cases are not fully independent; however both cases correspond to slightly different separations in UV magnitude. This allows us to explore in a limited way whether the observed separability signal shows any dependence on magnitude separation that is qualitatively consistent with trends seen in simulations. The uncertainties in the measurements are estimated using the Jackknife resampling method described in Section \ref{sec_Statistical Uncertainties}. The number of galaxies in each magnitude bin that make up the samples is tabulated in Table~\ref{table1}. 

It is clear from the figure that $s$ is below unity even for $\theta \lesssim 1500''$ (corresponding to scales as large as 11 Mpc) and the deviations increase with decreasing scale. The vertical line in each panel indicates an effective halo-exclusion scale, calculated as described in Section~\ref{sec_s_simul}, using the virial radius  given in Eq.~\eqref{eq_rvir}, with the average mass replaced by the effective mass of each observational sample. The procedure for computing the effective mass of observed sample is described later in this section.

At smaller angular scales, the deviation of $s$ from unity is consistently larger in the right panel of the figure. For instance, at $\theta \sim 100''$, $s$ reaches a value of 0.55 in the right panel, compared to approximately 0.62 in the left. Although this difference is not statistically significant given the current uncertainties, it may be indicative of a trend with increasing separation in UV magnitude between the bins. The scale dependence of $s$ is consistent with the trends seen in the $N$-body simulations presented in Section~\ref{sec_s_simul} and provides compelling evidence for the non-separable halo bias.

We quantify the deviation of $s$ from unity by performing the $\chi^2$ goodness-of-fit test, treating $s = 1$ as the null hypothesis (a constant function of $\theta$). For the left and right samples shown in Figure~\ref{fig_cf_ratio}, the $\chi^2$ values are $180.8$ and $346.5$, corresponding to p-values of $1.62 \times 10^{-30}$ and $1.09 \times 10^{-64}$, respectively. These large $\chi^2$ values and exceedingly small p-values strongly indicate that the parameter $s$ deviates significantly from unity in the observational data.

In order to compare the observational results for $s$ with those from $N$-body simulations, it is necessary to determine the halo masses corresponding to the galaxy samples used in the measurement. This is often done through a halo-model analysis of galaxy cross-correlations, which is beyond the scope of this work. Instead, we infer the effective halo mass of each observational sample from its measured large-scale bias. The separability function $s$ is then measured from $N$-body simulations for haloes with the same effective mass and compared directly with the observational results for $s$. 

To determine the effective mass of the galaxy samples defined by UV magnitude bins, we begin by estimating their large-scale bias using the method described in \citep{emy_jose_2024}. This requires computing the linear dark matter spatial correlation function, given by, 
\begin{equation}
    \xi^{\rm lin}_\text{mm}(r,z) = \dfrac{1}{2\pi^2}\int  dk \, P^{\rm lin}_{\rm mm}(k, z) \, \dfrac{\sin(kr)}{kr},  
\end{equation}
where $P^{\rm lin}_{\rm mm}$ is computed for Planck cosmology \citep{Planck_2018}. 

We applied the Limber transformation \citep{Limber1953} to $\xi^{\rm lin}_\text{mm}(r)$, using the redshift distribution of the sample galaxies given in Section~\ref{sec_hsc_ssp_data} to compute the linear dark matter angular correlation function $\omega^{\rm lin}_{\text{mm}}(\theta)$. The galaxy bias as a function of $\theta$ is then estimated for each magnitude-binned sample using its observed angular auto-correlation function as 
\begin{equation}
   b_g^2(\theta) = \omega_{\rm a}(\theta)/\omega^{\rm lin}_{\text{mm}}(\theta). 
\end{equation}
As in \citet{emy_jose_2024}, we find that the bias remains approximately constant over the angular range $100'' \leq \theta \leq 500''$ for all galaxy samples. We therefore define the effective large-scale galaxy bias of the sample as: 
\begin{equation}
    b^2_\text{eff} = \dfrac{\int \omega_a(\theta) \, \theta^2 \, d\theta } {\int \omega^{\rm lin}_{\text{mm}}(\theta) \, \theta^2 \, d\theta},
\end{equation}
where the integrals are evaluated over the angular range $\theta = 100''$ to $500''$. The resulting effective bias, $b_{\rm eff}$ is listed in Table~\ref{table1} for each sample. The errors in the measured galaxy angular correlation functions, determined using the Jackknife method, are propagated to estimate the uncertainties in the effective bias (also given in Table~\ref{table1}).

The relation between halo mass and bias (discussed in Section~\ref{sec_acf_ccf}) has been well calibrated using $N$-body simulations (e.g., \citealt{sheth_tormen_1999_bias, Tinker+2010_bias}). We determine the effective halo mass of the sample, $\rm M_\text{eff}$, from its effective large-scale bias, $\rm b_\text{eff}$, by applying the halo bias–mass relation of \citet{Tinker+2010_bias}, propagating the uncertainty in $\rm b_\text{eff}$ to estimate the error in $\rm M_\text{eff}$. Our estimates of the $\rm M_\text{eff}$ of the UV magnitude-binned galaxy samples ($[\rm m_\text{uv}]$) used to measure $s$, along with their associated uncertainties, are listed in Table~\ref{table1}. It is worth noting that this effective halo mass may not precisely correspond to the average halo mass of the sample, as derived from a full halo model analysis; nonetheless, it provides a robust and useful approximation.

We next measure $s$ corresponding to the observed galaxy samples directly from simulations. Since $N$-body simulation data are unavailable at the observed redshift of $z = 3.6$, we instead use data at $z \sim 3$, the closest available snapshot. To do this, we select haloes from the simulation in mass bins matching the 1-$\sigma$ range of $\rm M_\text{eff}$ for each of the two magnitude-binned galaxy samples used in the observational measurement of $s$. For example, the halo mass bin corresponding to the sample with $20 \leq \rm m_{\rm UV} < 23.9$ and $\rm M_{\rm eff} = 9.86\substack{+2.1 \\ -1.9}\times 10^{12} \text M_\odot$ is defined as $7.96 \times 10^{12} \leq M/\text M_\odot < 1.196 \times 10^{13}$. 
We compute the spatial halo auto- and cross-correlation functions of these samples as described in the previous section. These spatial correlation functions are then converted into angular correlation functions using the Limber transformation, adopting the same redshift distribution $N(z)$ (in Fig.~\ref{fig_Nz}) as that of the observed galaxy sample. The resulting angular correlation functions are used to estimate the separability function, which is shown as blue circles in Fig.~\ref{fig_cf_ratio}. 

We find that the simulation-based estimates of $s$ slightly underpredict the observational measurements; however, the overall scale dependence and qualitative trends are in good agreement with observations. Moreover, the simulation snapshot used corresponds to a slightly lower redshift than the observations, where the separability signal is expected to be weaker. Taken together, this level of agreement suggests that the observed signal reflects genuine non-separability of halo bias. We emphasize that the present analysis is intended to provide an approximate comparison. A more accurate and self-consistent treatment, incorporating a full halo occupation distribution framework and detailed modeling of galaxy selection effects, is left for future work.

\subsection{Impact of satellite galaxies on the separability function}

The observed galaxy sample includes both central and satellite galaxies, whereas our simulation-based measurements of $s$ thus far have considered only dark matter haloes. Central galaxies reside at the centers of haloes, while satellites follow the dark matter distribution around the central galaxy. Therefore, measurements based solely on haloes primarily trace the clustering of central galaxies, without accounting for the contribution of satellites. To assess the impact of satellites, we populate haloes with both the central and the satellite galaxies and re-measure $s$. This enables us to evaluate whether the observed bias separability function remains robust in the presence of satellite galaxies. 

To populate dark matter haloes with galaxies, one requires a Halo Occupation Distribution (HOD) model \citep{cooray_sheth_2002}, which provides a statistical framework for describing the number of central and satellite galaxies residing in haloes of a given mass. We employ a simple HOD prescription to generate a mock galaxy catalogue from the $N$-body simulations that reflects the observed HSC-SSP galaxy population.  For this we have utilized the Abacus Summit huge simulations with a box size of $7500$ Mpc/h. 

We begin by assuming a simple scaling relation in which the luminosity of a central galaxy is proportional to its host halo mass ($L = kM$, where $k$ is a constant). Under this assumption, the mean number of central galaxies, $\bar{N}_c$, in a sample defined by a luminosity threshold $L_{\rm th}$ follows a step function:
\begin{equation}
    \bar{N_c}=\Theta(M,M_{\rm th}), 
\end{equation}
where $L_{\rm th} = k M_{\rm th}$. In this model, all dark matter haloes with mass exceeding $M_{\rm th}$ host a central galaxy with luminosity greater than $L_{\rm th}$, located at the halo center and assigned the same mass as the parent halo \citep{Tinker_2005_bias}.

The average number of satellite galaxies in a halo of mass $M$, $\bar{N_s}$, is modelled as a power law \citep{harikane2022goldrush}:  
\begin{equation}
\bar{N_s} = 
\begin{cases}
 \left(\dfrac{M-M_{\rm th}}{M_1}\right)^\alpha & \text{if } M \geq M_{\rm th}, \\
  0  & \text{if } M < M_{\rm th}. 
\end{cases}
\end{equation}
Following \citet{harikane2022goldrush}, we adopt $\alpha = 1$. The number of satellite galaxies within a dark matter halo of mass $M$ is drawn from a Poisson distribution with mean $\bar{N}_s$. The masses of the satellite galaxies brighter than the luminosity threshold $L_{\rm th}$ are assigned using the subhalo mass function (SHMF), which describes the average number of satellites of mass $M_s$ residing within a halo of mass $M$:
\begin{equation}
{\rm SHMF}(M_{\rm s} \mid M) = 
\begin{cases}
 A \left(\frac{M_{\rm s}}{\gamma M}\right)^{-\beta}, \quad  &M_{\rm th} \leq  M_{\rm s} <M/2, \\
 0 , &\text{Otherwise }. 
\end{cases}
\end{equation}
where we adopt $\beta = 2$ and $\gamma = 0.5$, following \citet{lee+2009} (see also \citet{bosch2004mass,jiang2016statistics}). Thus, the maximum mass of a satellite galaxy is limited to half the mass of its host halo. The constant \( A \) is fixed using the average number of satellites, \( N_s(M) \). The luminosity of a satellite galaxy is also assumed to be $L_s = k M_{\rm s}$. The spatial distribution of satellite galaxies around the central galaxy is modeled using random positions drawn from a Navarro–Frenk–White profile. In the absence of satellites, the catalogue consists solely of central galaxies located at the centers of dark matter haloes with masses above the threshold $M_{\rm th}$.

The HOD parameters $M_{\rm th}$ and $M_1$ are chosen so that the resulting galaxy number density $n_g$ and satellite fraction $f_{\rm sat}$—defined as the ratio of the number density of satellite galaxies to the total galaxy number density and both computed directly from the simulations — match those of realistic galaxy samples used for measuring $s$.
For the catalogues constructed here, we use $M_{\rm th} =3.59 \times 10^{12} \text M_\odot$ and $M_1 =2.9 \times 10^{13} \text M_\odot$  to obtain a galaxy number density of $n_g = 5 \times 10^{-5} \, \mathrm{Mpc}^{-3}$ and a satellite fraction of $f_{\rm sat} = 5\%$.  For comparison, the number density of samples with a threshold luminosity of 24.5 is $6.1 \times 10^{-5}$, while the satellite fraction in the HSC-SSP survey is approximately 2\% (Table 8 of \citet{harikane2022goldrush}). 

Thus, the final catalogue contains both central and satellite galaxies with $M > M_{\rm th}$. In our model, galaxy luminosity is  proportional to halo mass with an arbitrary proportionality constant $k$. Therefore, instead of dividing the catalogue into luminosity/magnitude bins to measure $s$, we bin galaxies directly by mass. While more realistic HOD models treat luminosity as a stochastic function of halo mass, we leave such extensions to future work.

To assess the impact of satellite galaxies on $s$, we first split the full galaxy catalog—including both centrals and satellites—into two mass bins defined by $3 \times 10^{12} < M_1/M_\odot \leq 5 \times 10^{12}$  and $2.5 \times 10^{13}  < M_2/M_\odot \leq 5.5 \times 10^{13}$ , and measured $s([M_1], [M_2])$ following the procedure described in Section~\ref{sec_s_simul}. We then removed the satellites and recomputed $s$ for the same mass bins.

\begin{figure}
     \centering
           \includegraphics[width=0.92\linewidth]{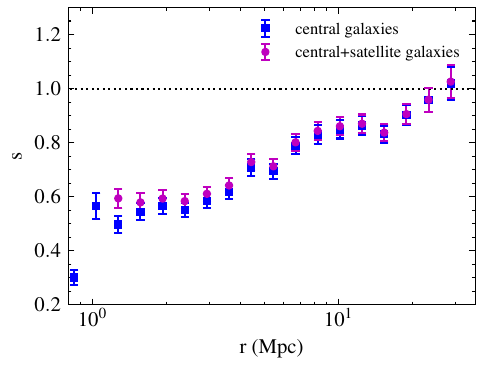}
           \caption{ The bias separability function with and without satellite galaxies, measured from a galaxy catalogue based on Abacus Summit Huge simulations.}
           \label{central+satellite gal}
 \end{figure}

The results, shown in Figure~\ref{central+satellite gal}, show that the presence of satellites slightly suppresses the deviation of $s$ from unity, thereby weakening the non-separability signal. Nevertheless, the signal remains clearly visible even with satellites included. This suggests that the observed non-separability is not an artifact introduced by satellite contamination, but a genuine feature of the galaxy distribution. The fact that the signal is even stronger in the absence of satellites further supports the robustness of our main result.

\section{Summary and Conclusion}
\label{sec_summary_conclusion}

In this work, we investigated a commonly used simplification in the halo model—the assumption of \textit{halo bias separability}, where the halo bias of two halo populations of different masses is expressed as a separable function of their masses (i.e. \(b^2_{\rm hh}(M_1, M_2) = b(M_1)\, b(M_2)\)). In particular, we examined the separability assumption using the Abacus dark matter \(N\)-body simulations covering the redshift range \(0 \leq z \leq 3\), and bright galaxy samples from the HSC-SSP Wide survey with a median redshift of \(\sim 3.6\).

We first defined the halo bias separability function, \(s(M_1, M_2, r, z)\), via Eq.~\eqref{eq_alpha} as the ratio of the cross-correlation function of haloes with masses \(M_1\) and \(M_2\) to the square root of the product of their auto-correlation functions. We then measured \(s\) from simulations using haloes spanning the mass range \(5 \times 10^{11}\)–\(5 \times 10^{13} \, \rm M_\odot\), typical of those hosting galaxies. The deviation of $s$ from unity directly quantifies the extent to which halo bias is non-separable. Our main findings from the simulations are summarised below:

\begin{itemize}
    \item \textbf{Scale dependence:} The separability function $s$ deviates from unity on quasi-linear scales ($\sim 1$--$5 \, \mathrm{Mpc}$), indicating a breakdown of bias separability at these scales. At larger separations, $s$ approaches unity, where halo bias is seperable.

    \item \textbf {Enhancement vs suppression:} Across all redshifts, the separability function \(s\) typically  lies below unity on quasi-linear scales, indicating a  reduced clustering of haloes of different masses. Notably, a small enhancement of \(s\) is observed at \(z \sim 0\) near \(1\,\mathrm{Mpc}\). 
    
    \item \textbf{Redshift dependence:} At $z \sim 0$, $s$ differs from unity by at most $\sim 10\%$, which increases to $\sim 20\%$ at $z \sim 1$. At $z \sim 3$, $s$ reaches a value as low as $0.5$, indicating a strong breakdown of the separability assumption at high redshifts and a corresponding suppression of halo cross-correlations relative to the  expectation from a separable case.

    \item \textbf{Mass dependence:} The deviation of $s$ from unity is larger for more massive haloes. It also increases with the mass separation between the halo mass bins used to measure $s$. At $z \sim 3$, the most massive haloes in our simulations show significant departures from unity even at scales of $\sim 15\,\mathrm{Mpc}$.
\end{itemize}

To test whether the non-separable bias is observable in galaxy clustering, we use galaxy samples from the HSC-SSP wide survey to measure the separability function at a median redshift of $z \sim 3.6$. Since the masses of individual galaxies are not directly available, we divide the galaxy sample into different UV magnitude bins and compute their angular auto- and cross-correlation functions. From these, we derive the separability function $s(\rm [m_{\text{uv1}}],\rm [m_{\text{uv2}}], \theta)$ using Eq.~\eqref{eq:s_def}. To our knowledge, this provides the first observational evidence for the non-separability of galaxy bias at high redshift, revealing the breakdown of a key halo model approximation in the early Universe. Specifically, we find that:

\begin{itemize}
    \item The measured $s$ deviates significantly from unity on angular scales $\theta \lesssim 1500''$. For example, for  samples defined by $20 <\rm m_\text{uv1} \leq 23.75$ and $24.1 <\rm m_\text{uv2} \leq 24.5$, $s$ reaches a value of 0.55 at $\theta \sim 100''$.
    
    \item The larger the separation in magnitude between galaxy samples, the greater the deviation of $s$ from unity. This is consistent with the mass-dependence of $s$ observed in simulations.
\end{itemize}

To compare the observed separability function with simulations, we estimate an effective halo mass for each galaxy sample using large-scale bias measurements, which is then used to compute $s$ from the simulations. The simulation-based estimates reproduce the overall scale dependence and trends seen in the observations. This agreement supports our conclusion that high-redshift galaxy bias is non-separable on quasi-linear scales.

To evaluate the effect of satellite galaxies on the measured separability function, we populated haloes from $N$-body simulations with both centrals and satellites using a simple HOD prescription. The HOD parameters are chosen such that the resulting galaxy number density and satellite fraction are in broad agreement with those of the observed HSC-SSP galaxy sample. Using this mock catalog, we measure $s$ with and without the inclusion of satellites. We find that satellites mildly suppress the deviation of $s$ from unity, but the non-separability signal remains significant in both cases. This suggests that the observed non-separability of halo bias is not a result of satellite contamination. Moreover, the fact that the signal is stronger in the absence of satellites supports the robustness of our findings.

Our results demonstrate that the assumption of halo bias separability breaks down on quasi-linear scales, particularly at high redshifts. This deviation is clearly detectable in observed galaxy clustering with current data. The strong agreement between simulations and observations lends robustness to this conclusion. These findings suggest that the separability function $s$ could serve as a powerful observational tool in upcoming surveys to constrain both galaxy formation physics and cosmological parameters. Future work using larger $N$-body simulations and improved analytic models will be crucial to fully exploring its potential.

\section*{Acknowledgements}
We thank the anonymous referee for a careful reading of the manuscript and for constructive comments that helped improve the clarity and presentation of this work.  EM acknowledges the financial support in the form of the fellowship of Cochin University of Science and Technology, Kerala. EM and Vipul acknowledges the financial support from Rashtriya Uchchatar Shiksha Abhiyan (RUSA 2.0). Suryan acknowledges the CSIR-SRF fellowship(File No:09/0239(11911)/2021-EMR-1).CJ is supported by the University Grant Commission, India through a BSR start-up grant (F.30-463/2019(BSR)) and the RUSA 2.0 scheme (No.CUSAT/PL(UGC).A1/2314/2023, No:T3A). CJ also acknowledges the access to the high-performance cluster at IUCAA (Pune, India) facilitated through the associateship program. We also acknowledge the use of ChatGPT for assistance in improving the clarity and grammar of the language in the manuscript. No scientific content, analysis, or interpretation was generated using this tool.

 \section*{Data Availability }
The N-body simulation data used in this article were accessed from \url{https://abacusnbody.org/}. The galaxy catalogues of the HSC-SSP survey were accessed from \url{https://hsc.mtk.nao.ac.jp/ssp/data-release/}. The derived data products in this article will be shared on reasonable request to the corresponding author.
 
\bibliographystyle{mnras}	

\bibliography{main}

@article{abacussummit_maksimova2021,
  title={AbacusSummit: a massive set of high-accuracy, high-resolution N-body simulations},
  author={Maksimova, Nina A and Garrison, Lehman H and Eisenstein, Daniel J and Hadzhiyska, Boryana and Bose, Sownak and Satterthwaite, Thomas P},
  journal={Monthly Notices of the Royal Astronomical Society},
  volume={508},
  number={3},
  pages={4017--4037},
  year={2021},
  publisher={Oxford University Press}
}

@article{abacus_garrison2021,
  title={The ABACUS cosmological N-body code},
  author={Garrison, Lehman H and Eisenstein, Daniel J and Ferrer, Douglas and Maksimova, Nina A and Pinto, Philip A},
  journal={Monthly Notices of the Royal Astronomical Society},
  volume={508},
  number={1},
  pages={575--596},
  year={2021},
  publisher={Oxford University Press}
}

@ARTICLE{aghanim+2018,
       author = {{Planck Collaboration} and {Aghanim}, N. and {Akrami}, Y. and {Ashdown}, M. and {Aumont}, J. and {Baccigalupi}, C. and {Ballardini}, M. and {Banday}, A.~J. and {Barreiro}, R.~B. and {Bartolo}, N. and {Basak}, S. and {Battye}, R. and {Benabed}, K. and {Bernard}, J. -P. and {Bersanelli}, M. and {Bielewicz}, P. and {Bock}, J.~J. and {Bond}, J.~R. and {Borrill}, J. and {Bouchet}, F.~R. and {Boulanger}, F. and {Bucher}, M. and {Burigana}, C. and {Butler}, R.~C. and {Calabrese}, E. and {Cardoso}, J. -F. and {Carron}, J. and {Challinor}, A. and {Chiang}, H.~C. and {Chluba}, J. and {Colombo}, L.~P.~L. and {Combet}, C. and {Contreras}, D. and {Crill}, B.~P. and {Cuttaia}, F. and {de Bernardis}, P. and {de Zotti}, G. and {Delabrouille}, J. and {Delouis}, J. -M. and {Di Valentino}, E. and {Diego}, J.~M. and {Dor{\'e}}, O. and {Douspis}, M. and {Ducout}, A. and {Dupac}, X. and {Dusini}, S. and {Efstathiou}, G. and {Elsner}, F. and {En{\ss}lin}, T.~A. and {Eriksen}, H.~K. and {Fantaye}, Y. and {Farhang}, M. and {Fergusson}, J. and {Fernandez-Cobos}, R. and {Finelli}, F. and {Forastieri}, F. and {Frailis}, M. and {Fraisse}, A.~A. and {Franceschi}, E. and {Frolov}, A. and {Galeotta}, S. and {Galli}, S. and {Ganga}, K. and {G{\'e}nova-Santos}, R.~T. and {Gerbino}, M. and {Ghosh}, T. and {Gonz{\'a}lez-Nuevo}, J. and {G{\'o}rski}, K.~M. and {Gratton}, S. and {Gruppuso}, A. and {Gudmundsson}, J.~E. and {Hamann}, J. and {Handley}, W. and {Hansen}, F.~K. and {Herranz}, D. and {Hildebrandt}, S.~R. and {Hivon}, E. and {Huang}, Z. and {Jaffe}, A.~H. and {Jones}, W.~C. and {Karakci}, A. and {Keih{\"a}nen}, E. and {Keskitalo}, R. and {Kiiveri}, K. and {Kim}, J. and {Kisner}, T.~S. and {Knox}, L. and {Krachmalnicoff}, N. and {Kunz}, M. and {Kurki-Suonio}, H. and {Lagache}, G. and {Lamarre}, J. -M. and {Lasenby}, A. and {Lattanzi}, M. and {Lawrence}, C.~R. and {Le Jeune}, M. and {Lemos}, P. and {Lesgourgues}, J. and {Levrier}, F. and {Lewis}, A. and {Liguori}, M. and {Lilje}, P.~B. and {Lilley}, M. and {Lindholm}, V. and {L{\'o}pez-Caniego}, M. and {Lubin}, P.~M. and {Ma}, Y. -Z. and {Mac{\'\i}as-P{\'e}rez}, J.~F. and {Maggio}, G. and {Maino}, D. and {Mandolesi}, N. and {Mangilli}, A. and {Marcos-Caballero}, A. and {Maris}, M. and {Martin}, P.~G. and {Martinelli}, M. and {Mart{\'\i}nez-Gonz{\'a}lez}, E. and {Matarrese}, S. and {Mauri}, N. and {McEwen}, J.~D. and {Meinhold}, P.~R. and {Melchiorri}, A. and {Mennella}, A. and {Migliaccio}, M. and {Millea}, M. and {Mitra}, S. and {Miville-Desch{\^e}nes}, M. -A. and {Molinari}, D. and {Montier}, L. and {Morgante}, G. and {Moss}, A. and {Natoli}, P. and {N{\o}rgaard-Nielsen}, H.~U. and {Pagano}, L. and {Paoletti}, D. and {Partridge}, B. and {Patanchon}, G. and {Peiris}, H.~V. and {Perrotta}, F. and {Pettorino}, V. and {Piacentini}, F. and {Polastri}, L. and {Polenta}, G. and {Puget}, J. -L. and {Rachen}, J.~P. and {Reinecke}, M. and {Remazeilles}, M. and {Renzi}, A. and {Rocha}, G. and {Rosset}, C. and {Roudier}, G. and {Rubi{\~n}o-Mart{\'\i}n}, J.~A. and {Ruiz-Granados}, B. and {Salvati}, L. and {Sandri}, M. and {Savelainen}, M. and {Scott}, D. and {Shellard}, E.~P.~S. and {Sirignano}, C. and {Sirri}, G. and {Spencer}, L.~D. and {Sunyaev}, R. and {Suur-Uski}, A. -S. and {Tauber}, J.~A. and {Tavagnacco}, D. and {Tenti}, M. and {Toffolatti}, L. and {Tomasi}, M. and {Trombetti}, T. and {Valenziano}, L. and {Valiviita}, J. and {Van Tent}, B. and {Vibert}, L. and {Vielva}, P. and {Villa}, F. and {Vittorio}, N. and {Wandelt}, B.~D. and {Wehus}, I.~K. and {White}, M. and {White}, S.~D.~M. and {Zacchei}, A. and {Zonca}, A.},
        title = "{Planck 2018 results. VI. Cosmological parameters}",
      journal = {\aap},
         year = 2020,
        month = sep,
       volume = {641},
          eid = {A6},
        pages = {A6},
          doi = {10.1051/0004-6361/201833910},
archivePrefix = {arXiv},
       eprint = {1807.06209},
 primaryClass = {astro-ph.CO},
       adsurl = {https://ui.adsabs.harvard.edu/abs/2020A&A...641A...6P}
}

@ARTICLE{abbott+2022,
       author = {{Abbott}, T.~M.~C. and {Aguena}, M. and {Alarcon}, A. and {Allam}, S. and {Alves}, O. and {Amon}, A. and {Andrade-Oliveira}, F. and {Annis}, J. and {Avila}, S. and {Bacon}, D. and {Baxter}, E. and {Bechtol}, K. and {Becker}, M.~R. and {Bernstein}, G.~M. and {Bhargava}, S. and {Birrer}, S. and {Blazek}, J. and {Brandao-Souza}, A. and {Bridle}, S.~L. and {Brooks}, D. and {Buckley-Geer}, E. and {Burke}, D.~L. and {Camacho}, H. and {Campos}, A. and {Carnero Rosell}, A. and {Carrasco Kind}, M. and {Carretero}, J. and {Castander}, F.~J. and {Cawthon}, R. and {Chang}, C. and {Chen}, A. and {Chen}, R. and {Choi}, A. and {Conselice}, C. and {Cordero}, J. and {Costanzi}, M. and {Crocce}, M. and {da Costa}, L.~N. and {da Silva Pereira}, M.~E. and {Davis}, C. and {Davis}, T.~M. and {De Vicente}, J. and {DeRose}, J. and {Desai}, S. and {Di Valentino}, E. and {Diehl}, H.~T. and {Dietrich}, J.~P. and {Dodelson}, S. and {Doel}, P. and {Doux}, C. and {Drlica-Wagner}, A. and {Eckert}, K. and {Eifler}, T.~F. and {Elsner}, F. and {Elvin-Poole}, J. and {Everett}, S. and {Evrard}, A.~E. and {Fang}, X. and {Farahi}, A. and {Fernandez}, E. and {Ferrero}, I. and {Fert{\'e}}, A. and {Fosalba}, P. and {Friedrich}, O. and {Frieman}, J. and {Garc{\'\i}a-Bellido}, J. and {Gatti}, M. and {Gaztanaga}, E. and {Gerdes}, D.~W. and {Giannantonio}, T. and {Giannini}, G. and {Gruen}, D. and {Gruendl}, R.~A. and {Gschwend}, J. and {Gutierrez}, G. and {Harrison}, I. and {Hartley}, W.~G. and {Herner}, K. and {Hinton}, S.~R. and {Hollowood}, D.~L. and {Honscheid}, K. and {Hoyle}, B. and {Huff}, E.~M. and {Huterer}, D. and {Jain}, B. and {James}, D.~J. and {Jarvis}, M. and {Jeffrey}, N. and {Jeltema}, T. and {Kovacs}, A. and {Krause}, E. and {Kron}, R. and {Kuehn}, K. and {Kuropatkin}, N. and {Lahav}, O. and {Leget}, P. -F. and {Lemos}, P. and {Liddle}, A.~R. and {Lidman}, C. and {Lima}, M. and {Lin}, H. and {MacCrann}, N. and {Maia}, M.~A.~G. and {Marshall}, J.~L. and {Martini}, P. and {McCullough}, J. and {Melchior}, P. and {Mena-Fern{\'a}ndez}, J. and {Menanteau}, F. and {Miquel}, R. and {Mohr}, J.~J. and {Morgan}, R. and {Muir}, J. and {Myles}, J. and {Nadathur}, S. and {Navarro-Alsina}, A. and {Nichol}, R.~C. and {Ogando}, R.~L.~C. and {Omori}, Y. and {Palmese}, A. and {Pandey}, S. and {Park}, Y. and {Paz-Chinch{\'o}n}, F. and {Petravick}, D. and {Pieres}, A. and {Plazas Malag{\'o}n}, A.~A. and {Porredon}, A. and {Prat}, J. and {Raveri}, M. and {Rodriguez-Monroy}, M. and {Rollins}, R.~P. and {Romer}, A.~K. and {Roodman}, A. and {Rosenfeld}, R. and {Ross}, A.~J. and {Rykoff}, E.~S. and {Samuroff}, S. and {S{\'a}nchez}, C. and {Sanchez}, E. and {Sanchez}, J. and {Sanchez Cid}, D. and {Scarpine}, V. and {Schubnell}, M. and {Scolnic}, D. and {Secco}, L.~F. and {Serrano}, S. and {Sevilla-Noarbe}, I. and {Sheldon}, E. and {Shin}, T. and {Smith}, M. and {Soares-Santos}, M. and {Suchyta}, E. and {Swanson}, M.~E.~C. and {Tabbutt}, M. and {Tarle}, G. and {Thomas}, D. and {To}, C. and {Troja}, A. and {Troxel}, M.~A. and {Tucker}, D.~L. and {Tutusaus}, I. and {Varga}, T.~N. and {Walker}, A.~R. and {Weaverdyck}, N. and {Wechsler}, R. and {Weller}, J. and {Yanny}, B. and {Yin}, B. and {Zhang}, Y. and {Zuntz}, J. and {DES Collaboration}},
        title = "{Dark Energy Survey Year 3 results: Cosmological constraints from galaxy clustering and weak lensing}",
      journal = {\prd},
         year = 2022,
        month = jan,
       volume = {105},
       number = {2},
          eid = {023520},
        pages = {023520},
          doi = {10.1103/PhysRevD.105.023520},
archivePrefix = {arXiv},
       eprint = {2105.13549},
 primaryClass = {astro-ph.CO},
       adsurl = {https://ui.adsabs.harvard.edu/abs/2022PhRvD.105b3520A}
}

@ARTICLE{alam+2017,
       author = {{Alam}, Shadab and {Ata}, Metin and {Bailey}, Stephen and {Beutler}, Florian and {Bizyaev}, Dmitry and {Blazek}, Jonathan A. and {Bolton}, Adam S. and {Brownstein}, Joel R. and {Burden}, Angela and {Chuang}, Chia-Hsun and {Comparat}, Johan and {Cuesta}, Antonio J. and {Dawson}, Kyle S. and {Eisenstein}, Daniel J. and {Escoffier}, Stephanie and {Gil-Mar{\'\i}n}, H{\'e}ctor and {Grieb}, Jan Niklas and {Hand}, Nick and {Ho}, Shirley and {Kinemuchi}, Karen and {Kirkby}, David and {Kitaura}, Francisco and {Malanushenko}, Elena and {Malanushenko}, Viktor and {Maraston}, Claudia and {McBride}, Cameron K. and {Nichol}, Robert C. and {Olmstead}, Matthew D. and {Oravetz}, Daniel and {Padmanabhan}, Nikhil and {Palanque-Delabrouille}, Nathalie and {Pan}, Kaike and {Pellejero-Ibanez}, Marcos and {Percival}, Will J. and {Petitjean}, Patrick and {Prada}, Francisco and {Price-Whelan}, Adrian M. and {Reid}, Beth A. and {Rodr{\'\i}guez-Torres}, Sergio A. and {Roe}, Natalie A. and {Ross}, Ashley J. and {Ross}, Nicholas P. and {Rossi}, Graziano and {Rubi{\~n}o-Mart{\'\i}n}, Jose Alberto and {Saito}, Shun and {Salazar-Albornoz}, Salvador and {Samushia}, Lado and {S{\'a}nchez}, Ariel G. and {Satpathy}, Siddharth and {Schlegel}, David J. and {Schneider}, Donald P. and {Sc{\'o}ccola}, Claudia G. and {Seo}, Hee-Jong and {Sheldon}, Erin S. and {Simmons}, Audrey and {Slosar}, An{\v{z}}e and {Strauss}, Michael A. and {Swanson}, Molly E.~C. and {Thomas}, Daniel and {Tinker}, Jeremy L. and {Tojeiro}, Rita and {Maga{\~n}a}, Mariana Vargas and {Vazquez}, Jose Alberto and {Verde}, Licia and {Wake}, David A. and {Wang}, Yuting and {Weinberg}, David H. and {White}, Martin and {Wood-Vasey}, W. Michael and {Y{\`e}che}, Christophe and {Zehavi}, Idit and {Zhai}, Zhongxu and {Zhao}, Gong-Bo},
        title = "{The clustering of galaxies in the completed SDSS-III Baryon Oscillation Spectroscopic Survey: cosmological analysis of the DR12 galaxy sample}",
      journal = {\mnras},
         year = 2017,
        month = sep,
       volume = {470},
       number = {3},
        pages = {2617-2652},
          doi = {10.1093/mnras/stx721},
archivePrefix = {arXiv},
       eprint = {1607.03155},
 primaryClass = {astro-ph.CO},
       adsurl = {https://ui.adsabs.harvard.edu/abs/2017MNRAS.470.2617A}
}

@ARTICLE{aihara+2022_pdr3_subaru,
       author = {{Aihara}, Hiroaki and {AlSayyad}, Yusra and {Ando}, Makoto and {Armstrong}, Robert and {Bosch}, James and {Egami}, Eiichi and {Furusawa}, Hisanori and {Furusawa}, Junko and {Harasawa}, Sumiko and {Harikane}, Yuichi and {Hsieh}, Bau-Ching and {Ikeda}, Hiroyuki and {Ito}, Kei and {Iwata}, Ikuru and {Kodama}, Tadayuki and {Koike}, Michitaro and {Kokubo}, Mitsuru and {Komiyama}, Yutaka and {Li}, Xiangchong and {Liang}, Yongming and {Lin}, Yen-Ting and {Lupton}, Robert H. and {Lust}, Nate B. and {MacArthur}, Lauren A. and {Mawatari}, Ken and {Mineo}, Sogo and {Miyatake}, Hironao and {Miyazaki}, Satoshi and {More}, Surhud and {Morishima}, Takahiro and {Murayama}, Hitoshi and {Nakajima}, Kimihiko and {Nakata}, Fumiaki and {Nishizawa}, Atsushi J. and {Oguri}, Masamune and {Okabe}, Nobuhiro and {Okura}, Yuki and {Ono}, Yoshiaki and {Osato}, Ken and {Ouchi}, Masami and {Pan}, Yen-Chen and {Plazas Malag{\'o}n}, Andr{\'e}s A. and {Price}, Paul A. and {Reed}, Sophie L. and {Rykoff}, Eli S. and {Shibuya}, Takatoshi and {Simunovic}, Mirko and {Strauss}, Michael A. and {Sugimori}, Kanako and {Suto}, Yasushi and {Suzuki}, Nao and {Takada}, Masahiro and {Takagi}, Yuhei and {Takata}, Tadafumi and {Takita}, Satoshi and {Tanaka}, Masayuki and {Tang}, Shenli and {Taranu}, Dan S. and {Terai}, Tsuyoshi and {Toba}, Yoshiki and {Turner}, Edwin L. and {Uchiyama}, Hisakazu and {Vijarnwannaluk}, Bovornpratch and {Waters}, Christopher Z. and {Yamada}, Yoshihiko and {Yamamoto}, Naoaki and {Yamashita}, Takuji},
        title = "{Third data release of the Hyper Suprime-Cam Subaru Strategic Program}",
      journal = {\pasj},
         year = 2022,
        month = apr,
       volume = {74},
       number = {2},
        pages = {247-272},
          doi = {10.1093/pasj/psab122},
archivePrefix = {arXiv},
       eprint = {2108.13045},
 primaryClass = {astro-ph.IM},
       adsurl = {https://ui.adsabs.harvard.edu/abs/2022PASJ...74..247A}
}

@ARTICLE{aihara+2018_pdr1_subaru,
       author = {{Aihara}, Hiroaki and {Armstrong}, Robert and {Bickerton}, Steven and {Bosch}, James and {Coupon}, Jean and {Furusawa}, Hisanori and {Hayashi}, Yusuke and {Ikeda}, Hiroyuki and {Kamata}, Yukiko and {Karoji}, Hiroshi and {Kawanomoto}, Satoshi and {Koike}, Michitaro and {Komiyama}, Yutaka and {Lang}, Dustin and {Lupton}, Robert H. and {Mineo}, Sogo and {Miyatake}, Hironao and {Miyazaki}, Satoshi and {Morokuma}, Tomoki and {Obuchi}, Yoshiyuki and {Oishi}, Yukie and {Okura}, Yuki and {Price}, Paul A. and {Takata}, Tadafumi and {Tanaka}, Manobu M. and {Tanaka}, Masayuki and {Tanaka}, Yoko and {Uchida}, Tomohisa and {Uraguchi}, Fumihiro and {Utsumi}, Yousuke and {Wang}, Shiang-Yu and {Yamada}, Yoshihiko and {Yamanoi}, Hitomi and {Yasuda}, Naoki and {Arimoto}, Nobuo and {Chiba}, Masashi and {Finet}, Francois and {Fujimori}, Hiroki and {Fujimoto}, Seiji and {Furusawa}, Junko and {Goto}, Tomotsugu and {Goulding}, Andy and {Gunn}, James E. and {Harikane}, Yuichi and {Hattori}, Takashi and {Hayashi}, Masao and {He{\l}miniak}, Krzysztof G. and {Higuchi}, Ryo and {Hikage}, Chiaki and {Ho}, Paul T.~P. and {Hsieh}, Bau-Ching and {Huang}, Kuiyun and {Huang}, Song and {Imanishi}, Masatoshi and {Iwata}, Ikuru and {Jaelani}, Anton T. and {Jian}, Hung-Yu and {Kashikawa}, Nobunari and {Katayama}, Nobuhiko and {Kojima}, Takashi and {Konno}, Akira and {Koshida}, Shintaro and {Kusakabe}, Haruka and {Leauthaud}, Alexie and {Lee}, Chien-Hsiu and {Lin}, Lihwai and {Lin}, Yen-Ting and {Mandelbaum}, Rachel and {Matsuoka}, Yoshiki and {Medezinski}, Elinor and {Miyama}, Shoken and {Momose}, Rieko and {More}, Anupreeta and {More}, Surhud and {Mukae}, Shiro and {Murata}, Ryoma and {Murayama}, Hitoshi and {Nagao}, Tohru and {Nakata}, Fumiaki and {Niida}, Mana and {Niikura}, Hiroko and {Nishizawa}, Atsushi J. and {Oguri}, Masamune and {Okabe}, Nobuhiro and {Ono}, Yoshiaki and {Onodera}, Masato and {Onoue}, Masafusa and {Ouchi}, Masami and {Pyo}, Tae-Soo and {Shibuya}, Takatoshi and {Shimasaku}, Kazuhiro and {Simet}, Melanie and {Speagle}, Joshua and {Spergel}, David N. and {Strauss}, Michael A. and {Sugahara}, Yuma and {Sugiyama}, Naoshi and {Suto}, Yasushi and {Suzuki}, Nao and {Tait}, Philip J. and {Takada}, Masahiro and {Terai}, Tsuyoshi and {Toba}, Yoshiki and {Turner}, Edwin L. and {Uchiyama}, Hisakazu and {Umetsu}, Keiichi and {Urata}, Yuji and {Usuda}, Tomonori and {Yeh}, Sherry and {Yuma}, Suraphong},
        title = "{First data release of the Hyper Suprime-Cam Subaru Strategic Program}",
      journal = {\pasj},
         year = 2018,
        month = jan,
       volume = {70},
          eid = {S8},
        pages = {S8},
          doi = {10.1093/pasj/psx081},
archivePrefix = {arXiv},
       eprint = {1702.08449},
 primaryClass = {astro-ph.IM},
       adsurl = {https://ui.adsabs.harvard.edu/abs/2018PASJ...70S...8A}
}

@article{aihara2019second,
  title={Second data release of the Hyper Suprime-Cam Subaru strategic program},
  author={Aihara, Hiroaki and AlSayyad, Yusra and Ando, Makoto and Armstrong, Robert and Bosch, James and Egami, Eiichi and Furusawa, Hisanori and Furusawa, Junko and Goulding, Andy and Harikane, Yuichi and others},
  journal={Publications of the Astronomical Society of Japan},
  volume={71},
  number={6},
  pages={114},
  year={2019},
  publisher={Oxford University Press}
}

@article{barkana2007_nl_clustering,
    author = "Barkana, Rennan",
    title = "{On Correlated Random Walks and 21-cm Fluctuations During Cosmic Reionization}",
    eprint = "0704.3534",
    archivePrefix = "arXiv",
    primaryClass = "astro-ph",
    doi = "10.1111/j.1365-2966.2007.11569.x",
    journal = "Mon. Not. Roy. Astron. Soc.",
    volume = "376",
    pages = "1784--1792",
    year = "2007"
}

@ARTICLE{berti+2023_sham,
       author = {{Berti}, Angela M. and {Dawson}, Kyle S. and {Dominguez}, Wilber},
        title = "{The Galaxy-Halo Connection of DESI Luminous Red Galaxies with Subhalo Abundance Matching}",
      journal = {\apj},
         year = 2023,
        month = sep,
       volume = {954},
       number = {2},
          eid = {131},
        pages = {131},
          doi = {10.3847/1538-4357/ace76e},
archivePrefix = {arXiv},
       eprint = {2303.16096},
 primaryClass = {astro-ph.CO},
       adsurl = {https://ui.adsabs.harvard.edu/abs/2023ApJ...954..131B}
}

@ARTICLE{berlind_weinberg_2002,
       author = {{Berlind}, Andreas A. and {Weinberg}, David H.},
        title = "{The Halo Occupation Distribution: Toward an Empirical Determination of the Relation between Galaxies and Mass}",
      journal = {\apj},
         year = 2002,
        month = aug,
       volume = {575},
       number = {2},
        pages = {587-616},
          doi = {10.1086/341469},
archivePrefix = {arXiv},
       eprint = {astro-ph/0109001},
 primaryClass = {astro-ph},
       adsurl = {https://ui.adsabs.harvard.edu/abs/2002ApJ...575..587B}
}

@ARTICLE{Bosch+2018_hscpipe,
       author = {{Bosch}, James and {Armstrong}, Robert and {Bickerton}, Steven and {Furusawa}, Hisanori and {Ikeda}, Hiroyuki and {Koike}, Michitaro and {Lupton}, Robert and {Mineo}, Sogo and {Price}, Paul and {Takata}, Tadafumi and {Tanaka}, Masayuki and {Yasuda}, Naoki and {AlSayyad}, Yusra and {Becker}, Andrew C. and {Coulton}, William and {Coupon}, Jean and {Garmilla}, Jose and {Huang}, Song and {Krughoff}, K. Simon and {Lang}, Dustin and {Leauthaud}, Alexie and {Lim}, Kian-Tat and {Lust}, Nate B. and {MacArthur}, Lauren A. and {Mandelbaum}, Rachel and {Miyatake}, Hironao and {Miyazaki}, Satoshi and {Murata}, Ryoma and {More}, Surhud and {Okura}, Yuki and {Owen}, Russell and {Swinbank}, John D. and {Strauss}, Michael A. and {Yamada}, Yoshihiko and {Yamanoi}, Hitomi},
        title = "{The Hyper Suprime-Cam software pipeline}",
      journal = {\pasj},
         year = 2018,
        month = jan,
       volume = {70},
          eid = {S5},
        pages = {S5},
          doi = {10.1093/pasj/psx080},
archivePrefix = {arXiv},
       eprint = {1705.06766},
 primaryClass = {astro-ph.IM},
       adsurl = {https://ui.adsabs.harvard.edu/abs/2018PASJ...70S...5B}
}

@ARTICLE{bullock+2002,
       author = {{Bullock}, James S. and {Wechsler}, Risa H. and {Somerville}, Rachel S.},
        title = "{Galaxy halo occupation at high redshift}",
      journal = {\mnras},
         year = 2002,
        month = jan,
       volume = {329},
       number = {1},
        pages = {246-256},
          doi = {10.1046/j.1365-8711.2002.04959.x},
archivePrefix = {arXiv},
       eprint = {astro-ph/0106293},
 primaryClass = {astro-ph},
       adsurl = {https://ui.adsabs.harvard.edu/abs/2002MNRAS.329..246B}
}

@ARTICLE{bhowmick+2018,
       author = {{Bhowmick}, Aklant K. and {Di Matteo}, Tiziana and {Feng}, Yu and {Lanusse}, Francois},
        title = "{The clustering of z > 7 galaxies: predictions from the BLUETIDES simulation}",
      journal = {\mnras},
         year = 2018,
        month = mar,
       volume = {474},
       number = {4},
        pages = {5393-5405},
          doi = {10.1093/mnras/stx3149},
archivePrefix = {arXiv},
       eprint = {1707.02312},
 primaryClass = {astro-ph.CO},
       adsurl = {https://ui.adsabs.harvard.edu/abs/2018MNRAS.474.5393B}
}

@article{bosch2004mass,
  title={The Mass Function and Average Mass Loss Rate of Dark Matter Subhaloes},
  author={Bosch, Frank C and Tormen, Giuseppe and Giocoli, Carlo},
  journal={arXiv preprint astro-ph/0409201},
  year={2004}
}

@article{coupon2018bright,
  title={The bright-star masks for the HSC-SSP survey},
  author={Coupon, Jean and Czakon, Nicole and Bosch, James and Komiyama, Yutaka and Medezinski, Elinor and Miyazaki, Satoshi and Oguri, Masamune},
  journal={Publications of the Astronomical Society of Japan},
  volume={70},
  number={SP1},
  pages={S7},
  year={2018},
  publisher={Oxford University Press}
}

@ARTICLE{cooray_sheth_2002,
       author = {{Cooray}, Asantha and {Sheth}, Ravi},
        title = "{Halo models of large scale structure}",
      journal = {\physrep},
         year = 2002,
        month = dec,
       volume = {372},
       number = {1},
        pages = {1-129},
          doi = {10.1016/S0370-1573(02)00276-4},
archivePrefix = {arXiv},
       eprint = {astro-ph/0206508},
 primaryClass = {astro-ph},
       adsurl = {https://ui.adsabs.harvard.edu/abs/2002PhR...372....1C}
}

@ARTICLE{chaurasiya+2024,
       author = {{Chaurasiya}, Navin and {More}, Surhud and {Ishikawa}, Shogo and {Masaki}, Shogo and {Kashino}, Daichi and {Okumura}, Teppei},
        title = "{Galaxy-dark matter connection of photometric galaxies from the HSC-SSP Survey: galaxy-galaxy lensing and the halo model}",
      journal = {\mnras},
         year = 2024,
        month = jan,
       volume = {527},
       number = {3},
        pages = {5265-5292},
          doi = {10.1093/mnras/stad3340},
archivePrefix = {arXiv},
       eprint = {2307.03915},
 primaryClass = {astro-ph.GA},
       adsurl = {https://ui.adsabs.harvard.edu/abs/2024MNRAS.527.5265C}
}

@ARTICLE{conroy+2006,
       author = {{Conroy}, Charlie and {Wechsler}, Risa H. and {Kravtsov}, Andrey V.},
        title = "{Modeling Luminosity-dependent Galaxy Clustering through Cosmic Time}",
      journal = {\apj},
         year = 2006,
        month = aug,
       volume = {647},
       number = {1},
        pages = {201-214},
          doi = {10.1086/503602},
archivePrefix = {arXiv},
       eprint = {astro-ph/0512234},
 primaryClass = {astro-ph},
       adsurl = {https://ui.adsabs.harvard.edu/abs/2006ApJ...647..201C}
}

@article{dalmasso+2024_jwst,
    author = {Dalmasso, Nicolò and Leethochawalit, Nicha and Trenti, Michele and Boyett, Kristan},
    title = {Galaxy clustering at cosmic dawn from JWST/NIRCam observations to redshift z~11},
    journal = {Monthly Notices of the Royal Astronomical Society},
    volume = {533},
    number = {2},
    pages = {2391-2398},
    year = {2024},
    month = {08},
    issn = {0035-8711},
    doi = {10.1093/mnras/stae2006},
    url = {https://doi.org/10.1093/mnras/stae2006},
    eprint = {https://academic.oup.com/mnras/article-pdf/533/2/2391/58969026/stae2006.pdf},
}

@ARTICLE{desi_2016,
       author = {{DESI Collaboration} and {Aghamousa}, Amir and {Aguilar}, Jessica and {Ahlen}, Steve and {Alam}, Shadab and {Allen}, Lori E. and {Allende Prieto}, Carlos and {Annis}, James and {Bailey}, Stephen and {Balland}, Christophe and {Ballester}, Otger and {Baltay}, Charles and {Beaufore}, Lucas and {Bebek}, Chris and {Beers}, Timothy C. and {Bell}, Eric F. and {Bernal}, Jos{\'e} Luis and {Besuner}, Robert and {Beutler}, Florian and {Blake}, Chris and {Bleuler}, Hannes and {Blomqvist}, Michael and {Blum}, Robert and {Bolton}, Adam S. and {Briceno}, Cesar and {Brooks}, David and {Brownstein}, Joel R. and {Buckley-Geer}, Elizabeth and {Burden}, Angela and {Burtin}, Etienne and {Busca}, Nicolas G. and {Cahn}, Robert N. and {Cai}, Yan-Chuan and {Cardiel-Sas}, Laia and {Carlberg}, Raymond G. and {Carton}, Pierre-Henri and {Casas}, Ricard and {Castander}, Francisco J. and {Cervantes-Cota}, Jorge L. and {Claybaugh}, Todd M. and {Close}, Madeline and {Coker}, Carl T. and {Cole}, Shaun and {Comparat}, Johan and {Cooper}, Andrew P. and {Cousinou}, M. -C. and {Crocce}, Martin and {Cuby}, Jean-Gabriel and {Cunningham}, Daniel P. and {Davis}, Tamara M. and {Dawson}, Kyle S. and {de la Macorra}, Axel and {De Vicente}, Juan and {Delubac}, Timoth{\'e}e and {Derwent}, Mark and {Dey}, Arjun and {Dhungana}, Govinda and {Ding}, Zhejie and {Doel}, Peter and {Duan}, Yutong T. and {Ealet}, Anne and {Edelstein}, Jerry and {Eftekharzadeh}, Sarah and {Eisenstein}, Daniel J. and {Elliott}, Ann and {Escoffier}, St{\'e}phanie and {Evatt}, Matthew and {Fagrelius}, Parker and {Fan}, Xiaohui and {Fanning}, Kevin and {Farahi}, Arya and {Farihi}, Jay and {Favole}, Ginevra and {Feng}, Yu and {Fernandez}, Enrique and {Findlay}, Joseph R. and {Finkbeiner}, Douglas P. and {Fitzpatrick}, Michael J. and {Flaugher}, Brenna and {Flender}, Samuel and {Font-Ribera}, Andreu and {Forero-Romero}, Jaime E. and {Fosalba}, Pablo and {Frenk}, Carlos S. and {Fumagalli}, Michele and {Gaensicke}, Boris T. and {Gallo}, Giuseppe and {Garcia-Bellido}, Juan and {Gaztanaga}, Enrique and {Pietro Gentile Fusillo}, Nicola and {Gerard}, Terry and {Gershkovich}, Irena and {Giannantonio}, Tommaso and {Gillet}, Denis and {Gonzalez-de-Rivera}, Guillermo and {Gonzalez-Perez}, Violeta and {Gott}, Shelby and {Graur}, Or and {Gutierrez}, Gaston and {Guy}, Julien and {Habib}, Salman and {Heetderks}, Henry and {Heetderks}, Ian and {Heitmann}, Katrin and {Hellwing}, Wojciech A. and {Herrera}, David A. and {Ho}, Shirley and {Holland}, Stephen and {Honscheid}, Klaus and {Huff}, Eric and {Hutchinson}, Timothy A. and {Huterer}, Dragan and {Hwang}, Ho Seong and {Illa Laguna}, Joseph Maria and {Ishikawa}, Yuzo and {Jacobs}, Dianna and {Jeffrey}, Niall and {Jelinsky}, Patrick and {Jennings}, Elise and {Jiang}, Linhua and {Jimenez}, Jorge and {Johnson}, Jennifer and {Joyce}, Richard and {Jullo}, Eric and {Juneau}, St{\'e}phanie and {Kama}, Sami and {Karcher}, Armin and {Karkar}, Sonia and {Kehoe}, Robert and {Kennamer}, Noble and {Kent}, Stephen and {Kilbinger}, Martin and {Kim}, Alex G. and {Kirkby}, David and {Kisner}, Theodore and {Kitanidis}, Ellie and {Kneib}, Jean-Paul and {Koposov}, Sergey and {Kovacs}, Eve and {Koyama}, Kazuya and {Kremin}, Anthony and {Kron}, Richard and {Kronig}, Luzius and {Kueter-Young}, Andrea and {Lacey}, Cedric G. and {Lafever}, Robin and {Lahav}, Ofer and {Lambert}, Andrew and {Lampton}, Michael and {Landriau}, Martin and {Lang}, Dustin and {Lauer}, Tod R. and {Le Goff}, Jean-Marc and {Le Guillou}, Laurent and {Le Van Suu}, Auguste and {Lee}, Jae Hyeon and {Lee}, Su-Jeong and {Leitner}, Daniela and {Lesser}, Michael and {Levi}, Michael E. and {L'Huillier}, Benjamin and {Li}, Baojiu and {Liang}, Ming and {Lin}, Huan and {Linder}, Eric and {Loebman}, Sarah R. and {Luki{\'c}}, Zarija and {Ma}, Jun and {MacCrann}, Niall and {Magneville}, Christophe and {Makarem}, Laleh and {Manera}, Marc and {Manser}, Christopher J. and {Marshall}, Robert and {Martini}, Paul and {Massey}, Richard and {Matheson}, Thomas and {McCauley}, Jeremy and {McDonald}, Patrick and {McGreer}, Ian D. and {Meisner}, Aaron and {Metcalfe}, Nigel and {Miller}, Timothy N. and {Miquel}, Ramon and {Moustakas}, John and {Myers}, Adam and {Naik}, Milind and {Newman}, Jeffrey A. and {Nichol}, Robert C. and {Nicola}, Andrina and {Nicolati da Costa}, Luiz and {Nie}, Jundan and {Niz}, Gustavo and {Norberg}, Peder and {Nord}, Brian and {Norman}, Dara and {Nugent}, Peter and {O'Brien}, Thomas and {Oh}, Minji and {Olsen}, Knut A.~G.},
        title = "{The DESI Experiment Part I: Science,Targeting, and Survey Design}",
      journal = {arXiv e-prints},
         year = 2016,
        month = oct,
          eid = {arXiv:1611.00036},
        pages = {arXiv:1611.00036},
          doi = {10.48550/arXiv.1611.00036},
archivePrefix = {arXiv},
       eprint = {1611.00036},
 primaryClass = {astro-ph.IM},
       adsurl = {https://ui.adsabs.harvard.edu/abs/2016arXiv161100036D}
}

@ARTICLE{dvornik+2023,
       author = {{Dvornik}, Andrej and {Heymans}, Catherine and {Asgari}, Marika and {Mahony}, Constance and {Joachimi}, Benjamin and {Bilicki}, Maciej and {Chisari}, Elisa and {Hildebrandt}, Hendrik and {Hoekstra}, Henk and {Johnston}, Harry and {Kuijken}, Konrad and {Mead}, Alexander and {Miyatake}, Hironao and {Nishimichi}, Takahiro and {Reischke}, Robert and {Unruh}, Sandra and {Wright}, Angus H.},
        title = "{KiDS-1000: Combined halo-model cosmology constraints from galaxy abundance, galaxy clustering, and galaxy-galaxy lensing}",
      journal = {\aap},
         year = 2023,
        month = jul,
       volume = {675},
          eid = {A189},
        pages = {A189},
          doi = {10.1051/0004-6361/202245158},
archivePrefix = {arXiv},
       eprint = {2210.03110},
 primaryClass = {astro-ph.CO},
       adsurl = {https://ui.adsabs.harvard.edu/abs/2023A&A...675A.189D}
}

@article{emy_jose_2024,
  title={Probing Environmental Dependence of High-Redshift Galaxy Properties with the Marked Correlation Function},
  author={Mons, Emy and Jose, Charles},
  journal={arXiv preprint arXiv:2412.12573},
  year={2024}
}

@ARTICLE{ euclid_2020_cosmology,
       author = {{Euclid Collaboration} and {Blanchard}, A. and {Camera}, S. and {Carbone}, C. and {Cardone}, V.~F. and {Casas}, S. and {Clesse}, S. and {Ili{\'c}}, S. and {Kilbinger}, M. and {Kitching}, T. and {Kunz}, M. and {Lacasa}, F. and {Linder}, E. and {Majerotto}, E. and {Markovi{\v{c}}}, K. and {Martinelli}, M. and {Pettorino}, V. and {Pourtsidou}, A. and {Sakr}, Z. and {S{\'a}nchez}, A.~G. and {Sapone}, D. and {Tutusaus}, I. and {Yahia-Cherif}, S. and {Yankelevich}, V. and {Andreon}, S. and {Aussel}, H. and {Balaguera-Antol{\'\i}nez}, A. and {Baldi}, M. and {Bardelli}, S. and {Bender}, R. and {Biviano}, A. and {Bonino}, D. and {Boucaud}, A. and {Bozzo}, E. and {Branchini}, E. and {Brau-Nogue}, S. and {Brescia}, M. and {Brinchmann}, J. and {Burigana}, C. and {Cabanac}, R. and {Capobianco}, V. and {Cappi}, A. and {Carretero}, J. and {Carvalho}, C.~S. and {Casas}, R. and {Castander}, F.~J. and {Castellano}, M. and {Cavuoti}, S. and {Cimatti}, A. and {Cledassou}, R. and {Colodro-Conde}, C. and {Congedo}, G. and {Conselice}, C.~J. and {Conversi}, L. and {Copin}, Y. and {Corcione}, L. and {Coupon}, J. and {Courtois}, H.~M. and {Cropper}, M. and {Da Silva}, A. and {de la Torre}, S. and {Di Ferdinando}, D. and {Dubath}, F. and {Ducret}, F. and {Duncan}, C.~A.~J. and {Dupac}, X. and {Dusini}, S. and {Fabbian}, G. and {Fabricius}, M. and {Farrens}, S. and {Fosalba}, P. and {Fotopoulou}, S. and {Fourmanoit}, N. and {Frailis}, M. and {Franceschi}, E. and {Franzetti}, P. and {Fumana}, M. and {Galeotta}, S. and {Gillard}, W. and {Gillis}, B. and {Giocoli}, C. and {G{\'o}mez-Alvarez}, P. and {Graci{\'a}-Carpio}, J. and {Grupp}, F. and {Guzzo}, L. and {Hoekstra}, H. and {Hormuth}, F. and {Israel}, H. and {Jahnke}, K. and {Keihanen}, E. and {Kermiche}, S. and {Kirkpatrick}, C.~C. and {Kohley}, R. and {Kubik}, B. and {Kurki-Suonio}, H. and {Ligori}, S. and {Lilje}, P.~B. and {Lloro}, I. and {Maino}, D. and {Maiorano}, E. and {Marggraf}, O. and {Martinet}, N. and {Marulli}, F. and {Massey}, R. and {Medinaceli}, E. and {Mei}, S. and {Mellier}, Y. and {Metcalf}, B. and {Metge}, J.~J. and {Meylan}, G. and {Moresco}, M. and {Moscardini}, L. and {Munari}, E. and {Nichol}, R.~C. and {Niemi}, S. and {Nucita}, A.~A. and {Padilla}, C. and {Paltani}, S. and {Pasian}, F. and {Percival}, W.~J. and {Pires}, S. and {Polenta}, G. and {Poncet}, M. and {Pozzetti}, L. and {Racca}, G.~D. and {Raison}, F. and {Renzi}, A. and {Rhodes}, J. and {Romelli}, E. and {Roncarelli}, M. and {Rossetti}, E. and {Saglia}, R. and {Schneider}, P. and {Scottez}, V. and {Secroun}, A. and {Sirri}, G. and {Stanco}, L. and {Starck}, J. -L. and {Sureau}, F. and {Tallada-Cresp{\'\i}}, P. and {Tavagnacco}, D. and {Taylor}, A.~N. and {Tenti}, M. and {Tereno}, I. and {Toledo-Moreo}, R. and {Torradeflot}, F. and {Valenziano}, L. and {Vassallo}, T. and {Verdoes Kleijn}, G.~A. and {Viel}, M. and {Wang}, Y. and {Zacchei}, A. and {Zoubian}, J. and {Zucca}, E.},
        title = "{Euclid preparation. VII. Forecast validation for Euclid cosmological probes}",
      journal = {\aap},
         year = 2020,
        month = oct,
       volume = {642},
          eid = {A191},
        pages = {A191},
          doi = {10.1051/0004-6361/202038071},
archivePrefix = {arXiv},
       eprint = {1910.09273},
 primaryClass = {astro-ph.CO},
       adsurl = {https://ui.adsabs.harvard.edu/abs/2020A&A...642A.191E}
}

@ARTICLE{giavalisco+2001,
       author = {{Giavalisco}, Mauro and {Dickinson}, Mark},
        title = "{Clustering Segregation with Ultraviolet Luminosity in Lyman Break Galaxies at z\raisebox{-0.5ex}\textasciitilde3 and Its Implications}",
      journal = {\apj},
         year = 2001,
        month = mar,
       volume = {550},
       number = {1},
        pages = {177-194},
          doi = {10.1086/319715},
archivePrefix = {arXiv},
       eprint = {astro-ph/0012249},
 primaryClass = {astro-ph},
       adsurl = {https://ui.adsabs.harvard.edu/abs/2001ApJ...550..177G}
}

@ARTICLE{gsponer+2024,
       author = {{Gsponer}, Rafaela and {Zhao}, Ruiyang and {Donald-McCann}, Jamie and {Bacon}, David and {Koyama}, Kazuya and {Crittenden}, Robert and {Simon}, Th{\'e}o and {Mueller}, Eva-Maria},
        title = "{Cosmological constraints on early dark energy from the full shape analysis of eBOSS DR16}",
      journal = {\mnras},
         year = 2024,
        month = may,
       volume = {530},
       number = {3},
        pages = {3075-3099},
          doi = {10.1093/mnras/stae992},
archivePrefix = {arXiv},
       eprint = {2312.01977},
 primaryClass = {astro-ph.CO},
       adsurl = {https://ui.adsabs.harvard.edu/abs/2024MNRAS.530.3075G}
}

@article{giavalisco2002lyman,
  title={Lyman-break galaxies},
  author={Giavalisco, Mauro},
  journal={Annual Review of Astronomy and Astrophysics},
  volume={40},
  number={1},
  pages={579--641},
  year={2002},
  publisher={Annual Reviews 4139 El Camino Way, PO Box 10139, Palo Alto, CA 94303-0139, USA}
}

@ARTICLE{gao_2023_lrg_elg_DESI,
       author = {{Gao}, Hongyu and {Jing}, Y.~P. and {Gui}, Shanquan and {Xu}, Kun and {Zheng}, Yun and {Zhao}, Donghai and {Aguilar}, Jessica Nicole and {Ahlen}, Steven and {Brooks}, David and {Claybaugh}, Todd and {Dawson}, Kyle and {xde la Macorra}, Axel and {Doel}, Peter and {Fanning}, Kevin and {Forero-Romero}, Jaime E. and {A Gontcho}, Satya Gontcho and {Guy}, Julien and {Honscheid}, Klaus and {Kehoe}, Robert and {Landriau}, Martin and {Manera}, Marc and {Meisner}, Aaron and {Miquel}, Ramon and {Moustakas}, John and {Newman}, Jeffrey A. and {Nie}, Jundan and {Percival}, Will and {Rossi}, Graziano and {Schubnell}, Michael and {Seo}, Hee-Jong and {Tarl{\'e}}, Gregory and {Weaver}, Benjamin Alan and {Yu}, Jiaxi and {Zhou}, Zhimin},
        title = "{The DESI One-Percent Survey: Constructing Galaxy-Halo Connections for ELGs and LRGs Using Auto and Cross Correlations}",
      journal = {\apj},
         year = 2023,
        month = sep,
       volume = {954},
       number = {2},
          eid = {207},
        pages = {207},
          doi = {10.3847/1538-4357/ace90a},
archivePrefix = {arXiv},
       eprint = {2306.06317},
 primaryClass = {astro-ph.GA},
       adsurl = {https://ui.adsabs.harvard.edu/abs/2023ApJ...954..207G}
}

@ARTICLE{hahn+2024,
       author = {{Hahn}, ChangHoon and {Lemos}, Pablo and {Parker}, Liam and {R{\'e}galdo-Saint Blancard}, Bruno and {Eickenberg}, Michael and {Ho}, Shirley and {Hou}, Jiamin and {Massara}, Elena and {Modi}, Chirag and {Moradinezhad Dizgah}, Azadeh and {Spergel}, David},
        title = "{Cosmological constraints from non-Gaussian and nonlinear galaxy clustering using the SIMBIG inference framework}",
      journal = {Nature Astronomy},
         year = 2024,
        month = nov,
       volume = {8},
        pages = {1457-1467},
          doi = {10.1038/s41550-024-02344-2},
       adsurl = {https://ui.adsabs.harvard.edu/abs/2024NatAs...8.1457H}
}

@article{harikane2018goldrush,
  title={GOLDRUSH. II. Clustering of galaxies at z~ 4--6 revealed with the half-million dropouts over the 100 deg2 area corresponding to 1 Gpc3},
  author={Harikane, Yuichi and Ouchi, Masami and Ono, Yoshiaki and Saito, Shun and Behroozi, Peter and More, Surhud and Shimasaku, Kazuhiro and Toshikawa, Jun and Lin, Yen-Ting and Akiyama, Masayuki and others},
  journal={Publications of the Astronomical Society of Japan},
  volume={70},
  number={SP1},
  pages={S11},
  year={2018},
  publisher={Oxford University Press}
}

@article{hsieh2014estimating,
  title={Estimating luminosities and stellar masses of galaxies photometrically without determining redshifts},
  author={Hsieh, BC and Yee, HKC},
  journal={The Astrophysical Journal},
  volume={792},
  number={2},
  pages={102},
  year={2014},
  publisher={IOP Publishing}
}

@article{hadzhiyska2022compaso,
  title={compaso: A new halo finder for competitive assignment to spherical overdensities},
  author={Hadzhiyska, Boryana and Eisenstein, Daniel and Bose, Sownak and Garrison, Lehman H and Maksimova, Nina},
  journal={Monthly Notices of the Royal Astronomical Society},
  volume={509},
  number={1},
  pages={501--521},
  year={2022},
  publisher={Oxford University Press}
}

@article{harikane2022goldrush,
  title={GOLDRUSH. IV. Luminosity Functions and Clustering Revealed with~ 4,000,000 Galaxies at z~ 2--7: Galaxy--AGN Transition, Star Formation Efficiency, and Implication for Evolution at z> 10},
  author={Harikane, Yuichi and Ono, Yoshiaki and Ouchi, Masami and Liu, Chengze and Sawicki, Marcin and Shibuya, Takatoshi and Behroozi, Peter S and He, Wanqiu and Shimasaku, Kazuhiro and Arnouts, Stephane and others},
  journal={The Astrophysical Journal Supplement Series},
  volume={259},
  number={1},
  pages={20},
  year={2022},
  publisher={IOP Publishing}
}

@ARTICLE{ivanov+2020,
       author = {{Ivanov}, Mikhail M. and {Simonovi{\'c}}, Marko and {Zaldarriaga}, Matias},
        title = "{Cosmological parameters from the BOSS galaxy power spectrum}",
      journal = {\jcap},
         year = 2020,
        month = may,
       volume = {2020},
       number = {5},
          eid = {042},
        pages = {042},
          doi = {10.1088/1475-7516/2020/05/042},
archivePrefix = {arXiv},
       eprint = {1909.05277},
 primaryClass = {astro-ph.CO},
       adsurl = {https://ui.adsabs.harvard.edu/abs/2020JCAP...05..042I}
}

@article{iliev+2003_nlclustering,
    author = "Iliev, Ilian T. and Scannapieco, Evan and Martel, Hugo and Shapiro, Paul R.",
    title = "{Nonlinear clustering during the cosmic Dark Ages and its effect on the 21-cm background from minihalos}",
    eprint = "astro-ph/0209216",
    archivePrefix = "arXiv",
    doi = "10.1046/j.1365-8711.2003.06410.x",
    journal = "Mon. Not. Roy. Astron. Soc.",
    volume = "341",
    pages = "81",
    year = "2003"
}

@ARTICLE{jose+2013,
       author = {{Jose}, Charles and {Subramanian}, Kandaswamy and {Srianand}, Raghunathan and {Samui}, Saumyadip},
        title = "{Spatial clustering of high-redshift Lyman-break galaxies}",
      journal = {\mnras},
         year = 2013,
        month = mar,
       volume = {429},
       number = {3},
        pages = {2333-2350},
          doi = {10.1093/mnras/sts503},
archivePrefix = {arXiv},
       eprint = {1208.2097},
 primaryClass = {astro-ph.CO},
       adsurl = {https://ui.adsabs.harvard.edu/abs/2013MNRAS.429.2333J}
}

@ARTICLE{jose+2014,
       author = {{Jose}, Charles and {Srianand}, Raghunathan and {Subramanian}, Kandaswamy},
        title = "{Clustering at high redshift: the connection between Lyman {\ensuremath{\alpha}} emitters and Lyman break galaxies}",
      journal = {\mnras},
         year = 2013,
        month = oct,
       volume = {435},
       number = {1},
        pages = {368-377},
          doi = {10.1093/mnras/stt1299},
archivePrefix = {arXiv},
       eprint = {1304.7458},
 primaryClass = {astro-ph.CO},
       adsurl = {https://ui.adsabs.harvard.edu/abs/2013MNRAS.435..368J}
}

@ARTICLE{jose+2016_nlbias,
       author = {{Jose}, Charles and {Lacey}, Cedric G. and {Baugh}, Carlton M.},
        title = "{The clustering of dark matter haloes: scale-dependent bias on quasi-linear scales}",
      journal = {\mnras},
         year = 2016,
        month = nov,
       volume = {463},
       number = {1},
        pages = {270-281},
          doi = {10.1093/mnras/stw1702},
archivePrefix = {arXiv},
       eprint = {1509.06715},
 primaryClass = {astro-ph.CO},
       adsurl = {https://ui.adsabs.harvard.edu/abs/2016MNRAS.463..270J}
}

@ARTICLE{jose+2017_lbg_acf,
       author = {{Jose}, Charles and {Baugh}, Carlton M. and {Lacey}, Cedric G. and {Subramanian}, Kandaswamy},
        title = "{Understanding the non-linear clustering of high-redshift galaxies}",
      journal = {\mnras},
         year = 2017,
        month = aug,
       volume = {469},
       number = {4},
        pages = {4428-4436},
          doi = {10.1093/mnras/stx1014},
archivePrefix = {arXiv},
       eprint = {1702.00853},
 primaryClass = {astro-ph.CO},
       adsurl = {https://ui.adsabs.harvard.edu/abs/2017MNRAS.469.4428J}
}

@ARTICLE{jimenez+2019,
       author = {{Jim{\'e}nez}, Esteban and {Contreras}, Sergio and {Padilla}, Nelson and {Zehavi}, Idit and {Baugh}, Carlton M. and {Gonzalez-Perez}, Violeta},
        title = "{Extensions to the halo occupation distribution model for more accurate clustering predictions}",
      journal = {\mnras},
         year = 2019,
        month = dec,
       volume = {490},
       number = {3},
        pages = {3532-3544},
          doi = {10.1093/mnras/stz2790},
archivePrefix = {arXiv},
       eprint = {1906.04298},
 primaryClass = {astro-ph.CO},
       adsurl = {https://ui.adsabs.harvard.edu/abs/2019MNRAS.490.3532J}
}

@article{jiang2016statistics,
  title={Statistics of dark matter substructure--I. Model and universal fitting functions},
  author={Jiang, Fangzhou and Van Den Bosch, Frank C},
  journal={Monthly Notices of the Royal Astronomical Society},
  volume={458},
  number={3},
  pages={2848--2869},
  year={2016},
  publisher={Oxford University Press}
}

@ARTICLE{kravtsov+2004,
       author = {{Kravtsov}, Andrey V. and {Berlind}, Andreas A. and {Wechsler}, Risa H. and {Klypin}, Anatoly A. and {Gottl{\"o}ber}, Stefan and {Allgood}, Brandon and {Primack}, Joel R.},
        title = "{The Dark Side of the Halo Occupation Distribution}",
      journal = {\apj},
         year = 2004,
        month = jul,
       volume = {609},
       number = {1},
        pages = {35-49},
          doi = {10.1086/420959},
archivePrefix = {arXiv},
       eprint = {astro-ph/0308519},
 primaryClass = {astro-ph},
       adsurl = {https://ui.adsabs.harvard.edu/abs/2004ApJ...609...35K}
}

@ARTICLE{kaiser_1983_bias,
       author = {{Kaiser}, N.},
        title = "{On the spatial correlations of Abell clusters.}",
      journal = {\apjl},
         year = 1984,
        month = sep,
       volume = {284},
        pages = {L9-L12},
          doi = {10.1086/184341},
       adsurl = {https://ui.adsabs.harvard.edu/abs/1984ApJ...284L...9K}
}

@article{kawanomoto2018hyper,
  title={Hyper Suprime-Cam: Filters},
  author={Kawanomoto, Satoshi and Uraguchi, Fumihiro and Komiyama, Yutaka and Miyazaki, Satoshi and Furusawa, Hisanori and Finet, Fran{\c{c}}ois and Hattori, Takashi and Wang, Shiang-Yu and Yasuda, Naoki and Suzuki, Naotaka},
  journal={Publications of the Astronomical Society of Japan},
  volume={70},
  number={4},
  pages={66},
  year={2018},
  publisher={Oxford University Press}
}

@ARTICLE{lap_2021_stochBias,
       author = {{Lapi}, Andrea and {Danese}, Luigi},
        title = "{A Stochastic Theory of the Hierarchical Clustering. II. Halo Progenitor Mass Function and Large-scale Bias}",
      journal = {\apj},
         year = 2021,
        month = apr,
       volume = {911},
       number = {1},
          eid = {11},
        pages = {11},
          doi = {10.3847/1538-4357/abe7eb},
archivePrefix = {arXiv},
       eprint = {2103.05279},
 primaryClass = {astro-ph.CO},
       adsurl = {https://ui.adsabs.harvard.edu/abs/2021ApJ...911...11L}
}

@article{landy1993bias,
  title={Bias and variance of angular correlation functions},
  author={Landy, Stephen D and Szalay, Alexander S},
  journal={Astrophysical Journal, Part 1 (ISSN 0004-637X), vol. 412, no. 1, p. 64-71.},
  volume={412},
  pages={64--71},
  year={1993}
}

@ARTICLE{lee+2009,
       author = {{Lee}, Kyoung-Soo and {Giavalisco}, Mauro and {Conroy}, Charlie and {Wechsler}, Risa H. and {Ferguson}, Henry C. and {Somerville}, Rachel S. and {Dickinson}, Mark E. and {Urry}, Claudia M.},
        title = "{Mapping the Dark Matter from UV Light at High Redshift: An Empirical Approach to Understand Galaxy Statistics}",
      journal = {\apj},
         year = 2009,
        month = apr,
       volume = {695},
       number = {1},
        pages = {368-390},
          doi = {10.1088/0004-637X/695/1/368},
archivePrefix = {arXiv},
       eprint = {0808.1727},
 primaryClass = {astro-ph},
       adsurl = {https://ui.adsabs.harvard.edu/abs/2009ApJ...695..368L}
}

@ARTICLE{Limber1953,
       author = {{Limber}, D. Nelson},
        title = "{The Analysis of Counts of the Extragalactic Nebulae in Terms of a Fluctuating Density Field.}",
      journal = {\apj},
         year = 1953,
        month = jan,
       volume = {117},
        pages = {134},
          doi = {10.1086/145672},
       adsurl = {https://ui.adsabs.harvard.edu/abs/1953ApJ...117..134L}
}

@ARTICLE{lsst_white_paper_212,
       author = {{LSST Dark Energy Science Collaboration}},
        title = "{Large Synoptic Survey Telescope: Dark Energy Science Collaboration}",
      journal = {arXiv e-prints},
         year = 2012,
        month = nov,
          eid = {arXiv:1211.0310},
        pages = {arXiv:1211.0310},
          doi = {10.48550/arXiv.1211.0310},
archivePrefix = {arXiv},
       eprint = {1211.0310},
 primaryClass = {astro-ph.CO},
       adsurl = {https://ui.adsabs.harvard.edu/abs/2012arXiv1211.0310L}
}

@ARTICLE{masaki_2014_lrg,
       author = {{Masaki}, Shogo and {Hikage}, Chiaki and {Takada}, Masahiro and {Spergel}, David N. and {Sugiyama}, Naoshi},
        title = "{Understanding the nature of luminous red galaxies (LRGs): connecting LRGs to central and satellite subhaloes}",
      journal = {\mnras},
         year = 2013,
        month = aug,
       volume = {433},
       number = {4},
        pages = {3506-3522},
          doi = {10.1093/mnras/stt981},
archivePrefix = {arXiv},
       eprint = {1211.7077},
 primaryClass = {astro-ph.CO},
       adsurl = {https://ui.adsabs.harvard.edu/abs/2013MNRAS.433.3506M}
}

@ARTICLE{mahony+2022_blb_halomodel,
       author = {{Mahony}, Constance and {Dvornik}, Andrej and {Mead}, Alexander and {Heymans}, Catherine and {Asgari}, Marika and {Hildebrandt}, Hendrik and {Miyatake}, Hironao and {Nishimichi}, Takahiro and {Reischke}, Robert},
        title = "{The halo model with beyond-linear halo bias: unbiasing cosmological constraints from galaxy-galaxy lensing and clustering}",
      journal = {\mnras},
         year = 2022,
        month = sep,
       volume = {515},
       number = {2},
        pages = {2612-2623},
          doi = {10.1093/mnras/stac1858},
archivePrefix = {arXiv},
       eprint = {2202.01790},
 primaryClass = {astro-ph.CO},
       adsurl = {https://ui.adsabs.harvard.edu/abs/2022MNRAS.515.2612M}
}

@ARTICLE{mead_verde_2021,
       author = {{Mead}, A.~J. and {Verde}, L.},
        title = "{Including beyond-linear halo bias in halo models}",
      journal = {\mnras},
         year = 2021,
        month = may,
       volume = {503},
       number = {2},
        pages = {3095-3111},
          doi = {10.1093/mnras/stab748},
archivePrefix = {arXiv},
       eprint = {2011.08858},
 primaryClass = {astro-ph.CO},
       adsurl = {https://ui.adsabs.harvard.edu/abs/2021MNRAS.503.3095M}
}

@ARTICLE{moretti+2023,
       author = {{Moretti}, Chiara and {Tsedrik}, Maria and {Carrilho}, Pedro and {Pourtsidou}, Alkistis},
        title = "{Modified gravity and massive neutrinos: constraints from the full shape analysis of BOSS galaxies and forecasts for Stage IV surveys}",
      journal = {\jcap},
         year = 2023,
        month = dec,
       volume = {2023},
       number = {12},
          eid = {025},
        pages = {025},
          doi = {10.1088/1475-7516/2023/12/025},
archivePrefix = {arXiv},
       eprint = {2306.09275},
 primaryClass = {astro-ph.CO},
       adsurl = {https://ui.adsabs.harvard.edu/abs/2023JCAP...12..025M}
}

@ARTICLE{mo_white_1996,
       author = {{Mo}, H.~J. and {White}, S.~D.~M.},
        title = "{An analytic model for the spatial clustering of dark matter haloes}",
      journal = {\mnras},
         year = 1996,
        month = sep,
       volume = {282},
       number = {2},
        pages = {347-361},
          doi = {10.1093/mnras/282.2.347},
archivePrefix = {arXiv},
       eprint = {astro-ph/9512127},
 primaryClass = {astro-ph},
       adsurl = {https://ui.adsabs.harvard.edu/abs/1996MNRAS.282..347M}
}

@article{norberg2009statistical,
  title={Statistical analysis of galaxy surveys--I. Robust error estimation for two-point clustering statistics},
  author={Norberg, Peder and Baugh, Carlton M and Gaztanaga, Enrique and Croton, Darren J},
  journal={Monthly Notices of the Royal Astronomical Society},
  volume={396},
  number={1},
  pages={19--38},
  year={2009},
  publisher={Blackwell Publishing Ltd Oxford, UK}
}

@ARTICLE{okumura+2021,
       author = {{Okumura}, Teppei and {Hayashi}, Masao and {Chiu}, I. -Non and {Lin}, Yen-Ting and {Osato}, Ken and {Hsieh}, Bau-Ching and {Lin}, Sheng-Chieh},
        title = "{Angular clustering and host halo properties of [O II] emitters at z > 1 in the Subaru HSC survey}",
      journal = {\pasj},
         year = 2021,
        month = aug,
       volume = {73},
       number = {4},
        pages = {1186-1207},
          doi = {10.1093/pasj/psab068},
archivePrefix = {arXiv},
       eprint = {2012.12224},
 primaryClass = {astro-ph.GA},
       adsurl = {https://ui.adsabs.harvard.edu/abs/2021PASJ...73.1186O}
}

@article{ono2018great,
  title={Great Optically Luminous Dropout Research Using Subaru HSC (GOLDRUSH). I. UV luminosity functions at z~ 4--7 derived with the half-million dropouts on the 100 deg2 sky},
  author={Ono, Yoshiaki and Ouchi, Masami and Harikane, Yuichi and Toshikawa, Jun and Rauch, Michael and Yuma, Suraphong and Sawicki, Marcin and Shibuya, Takatoshi and Shimasaku, Kazuhiro and Oguri, Masamune and others},
  journal={Publications of the Astronomical Society of Japan},
  volume={70},
  number={SP1},
  pages={S10},
  year={2018},
  publisher={Oxford University Press}
}

@ARTICLE{park+2016,
       author = {{Park}, Jaehong and {Kim}, Han-Seek and {Wyithe}, J. Stuart B. and {Lacey}, C.~G. and {Baugh}, C.~M. and {Barone-Nugent}, R.~L. and {Trenti}, M. and {Bouwens}, R.~J.},
        title = "{The clustering and halo occupation distribution of Lyman-break galaxies at z {\ensuremath{\sim}} 4}",
      journal = {\mnras},
         year = 2016,
        month = sep,
       volume = {461},
       number = {1},
        pages = {176-189},
          doi = {10.1093/mnras/stw1316},
archivePrefix = {arXiv},
       eprint = {1511.01983},
 primaryClass = {astro-ph.GA},
       adsurl = {https://ui.adsabs.harvard.edu/abs/2016MNRAS.461..176P}
}

@ARTICLE{pellejero+2024,
       author = {{Pellejero Ib{\'a}{\~n}ez}, Marcos and {Angulo}, Raul E. and {Peacock}, John A.},
        title = "{Cosmological constraints from the full-shape galaxy power spectrum in SDSS-III BOSS using the BACCO hybrid Lagrangian bias emulator}",
      journal = {\mnras},
         year = 2024,
        month = nov,
       volume = {534},
       number = {4},
        pages = {3595-3611},
          doi = {10.1093/mnras/stae2319},
archivePrefix = {arXiv},
       eprint = {2407.07949},
 primaryClass = {astro-ph.CO},
       adsurl = {https://ui.adsabs.harvard.edu/abs/2024MNRAS.534.3595P}
}

@ARTICLE{pei+2024,
       author = {{Pei}, Wenxiang and {Guo}, Qi and {Li}, Ming and {Wang}, Qiao and {Han}, Jiaxin and {Hu}, Jia and {Su}, Tong and {Gao}, Liang and {Wang}, Jie and {Luo}, Yu and {Wei}, Chengliang},
        title = "{Simulating emission line galaxies for the next generation of large-scale structure surveys}",
      journal = {\mnras},
         year = 2024,
        month = apr,
       volume = {529},
       number = {4},
        pages = {4958-4979},
          doi = {10.1093/mnras/stae866},
archivePrefix = {arXiv},
       eprint = {2404.00092},
 primaryClass = {astro-ph.GA},
       adsurl = {https://ui.adsabs.harvard.edu/abs/2024MNRAS.529.4958P}
}

@book{Peebles1980,
    title={The large-scale structure of the universe},
  author={Peebles, Phillip James Edwin},
  year={2020},
  publisher={Princeton university press}

}

@ARTICLE{Planck_2018,
       author = {{Planck Collaboration} and {Aghanim}, N. and {Akrami}, Y. and {Ashdown}, M. and {Aumont}, J. and {Baccigalupi}, C. and {Ballardini}, M. and {Banday}, A.~J. and {Barreiro}, R.~B. and {Bartolo}, N. and {Basak}, S. and {Battye}, R. and {Benabed}, K. and {Bernard}, J. -P. and {Bersanelli}, M. and {Bielewicz}, P. and {Bock}, J.~J. and {Bond}, J.~R. and {Borrill}, J. and {Bouchet}, F.~R. and {Boulanger}, F. and {Bucher}, M. and {Burigana}, C. and {Butler}, R.~C. and {Calabrese}, E. and {Cardoso}, J. -F. and {Carron}, J. and {Challinor}, A. and {Chiang}, H.~C. and {Chluba}, J. and {Colombo}, L.~P.~L. and {Combet}, C. and {Contreras}, D. and {Crill}, B.~P. and {Cuttaia}, F. and {de Bernardis}, P. and {de Zotti}, G. and {Delabrouille}, J. and {Delouis}, J. -M. and {Di Valentino}, E. and {Diego}, J.~M. and {Dor{\'e}}, O. and {Douspis}, M. and {Ducout}, A. and {Dupac}, X. and {Dusini}, S. and {Efstathiou}, G. and {Elsner}, F. and {En{\ss}lin}, T.~A. and {Eriksen}, H.~K. and {Fantaye}, Y. and {Farhang}, M. and {Fergusson}, J. and {Fernandez-Cobos}, R. and {Finelli}, F. and {Forastieri}, F. and {Frailis}, M. and {Fraisse}, A.~A. and {Franceschi}, E. and {Frolov}, A. and {Galeotta}, S. and {Galli}, S. and {Ganga}, K. and {G{\'e}nova-Santos}, R.~T. and {Gerbino}, M. and {Ghosh}, T. and {Gonz{\'a}lez-Nuevo}, J. and {G{\'o}rski}, K.~M. and {Gratton}, S. and {Gruppuso}, A. and {Gudmundsson}, J.~E. and {Hamann}, J. and {Handley}, W. and {Hansen}, F.~K. and {Herranz}, D. and {Hildebrandt}, S.~R. and {Hivon}, E. and {Huang}, Z. and {Jaffe}, A.~H. and {Jones}, W.~C. and {Karakci}, A. and {Keih{\"a}nen}, E. and {Keskitalo}, R. and {Kiiveri}, K. and {Kim}, J. and {Kisner}, T.~S. and {Knox}, L. and {Krachmalnicoff}, N. and {Kunz}, M. and {Kurki-Suonio}, H. and {Lagache}, G. and {Lamarre}, J. -M. and {Lasenby}, A. and {Lattanzi}, M. and {Lawrence}, C.~R. and {Le Jeune}, M. and {Lemos}, P. and {Lesgourgues}, J. and {Levrier}, F. and {Lewis}, A. and {Liguori}, M. and {Lilje}, P.~B. and {Lilley}, M. and {Lindholm}, V. and {L{\'o}pez-Caniego}, M. and {Lubin}, P.~M. and {Ma}, Y. -Z. and {Mac{\'\i}as-P{\'e}rez}, J.~F. and {Maggio}, G. and {Maino}, D. and {Mandolesi}, N. and {Mangilli}, A. and {Marcos-Caballero}, A. and {Maris}, M. and {Martin}, P.~G. and {Martinelli}, M. and {Mart{\'\i}nez-Gonz{\'a}lez}, E. and {Matarrese}, S. and {Mauri}, N. and {McEwen}, J.~D. and {Meinhold}, P.~R. and {Melchiorri}, A. and {Mennella}, A. and {Migliaccio}, M. and {Millea}, M. and {Mitra}, S. and {Miville-Desch{\^e}nes}, M. -A. and {Molinari}, D. and {Montier}, L. and {Morgante}, G. and {Moss}, A. and {Natoli}, P. and {N{\o}rgaard-Nielsen}, H.~U. and {Pagano}, L. and {Paoletti}, D. and {Partridge}, B. and {Patanchon}, G. and {Peiris}, H.~V. and {Perrotta}, F. and {Pettorino}, V. and {Piacentini}, F. and {Polastri}, L. and {Polenta}, G. and {Puget}, J. -L. and {Rachen}, J.~P. and {Reinecke}, M. and {Remazeilles}, M. and {Renzi}, A. and {Rocha}, G. and {Rosset}, C. and {Roudier}, G. and {Rubi{\~n}o-Mart{\'\i}n}, J.~A. and {Ruiz-Granados}, B. and {Salvati}, L. and {Sandri}, M. and {Savelainen}, M. and {Scott}, D. and {Shellard}, E.~P.~S. and {Sirignano}, C. and {Sirri}, G. and {Spencer}, L.~D. and {Sunyaev}, R. and {Suur-Uski}, A. -S. and {Tauber}, J.~A. and {Tavagnacco}, D. and {Tenti}, M. and {Toffolatti}, L. and {Tomasi}, M. and {Trombetti}, T. and {Valenziano}, L. and {Valiviita}, J. and {Van Tent}, B. and {Vibert}, L. and {Vielva}, P. and {Villa}, F. and {Vittorio}, N. and {Wandelt}, B.~D. and {Wehus}, I.~K. and {White}, M. and {White}, S.~D.~M. and {Zacchei}, A. and {Zonca}, A.},
        title = "{Planck 2018 results. VI. Cosmological parameters}",
      journal = {\aap},
         year = 2020,
        month = sep,
       volume = {641},
          eid = {A6},
        pages = {A6},
          doi = {10.1051/0004-6361/201833910},
archivePrefix = {arXiv},
       eprint = {1807.06209},
 primaryClass = {astro-ph.CO},
       adsurl = {https://ui.adsabs.harvard.edu/abs/2020A&A...641A...6P}
}

@ARTICLE{pizzati_2024_qso_blb,
       author = {{Pizzati}, Elia and {Hennawi}, Joseph F. and {Schaye}, Joop and {Schaller}, Matthieu},
        title = "{Revisiting the extreme clustering of z {\ensuremath{\approx}} 4 quasars with large volume cosmological simulations}",
      journal = {\mnras},
         year = 2024,
        month = mar,
       volume = {528},
       number = {3},
        pages = {4466-4489},
          doi = {10.1093/mnras/stae329},
archivePrefix = {arXiv},
       eprint = {2311.17181},
 primaryClass = {astro-ph.GA},
       adsurl = {https://ui.adsabs.harvard.edu/abs/2024MNRAS.528.4466P}
}

@ARTICLE{reed+2007_nlbias,
       author = {{Reed}, Darren S. and {Bower}, Richard and {Frenk}, Carlos S. and {Jenkins}, Adrian and {Theuns}, Tom},
        title = "{The halo mass function from the dark ages through the present day}",
      journal = {\mnras},
         year = 2007,
        month = jan,
       volume = {374},
       number = {1},
        pages = {2-15},
          doi = {10.1111/j.1365-2966.2006.11204.x},
archivePrefix = {arXiv},
       eprint = {astro-ph/0607150},
 primaryClass = {astro-ph},
       adsurl = {https://ui.adsabs.harvard.edu/abs/2007MNRAS.374....2R}
}

@article{steidel1996spectroscopic,
  title={Spectroscopic confirmation of a population of normal star-forming galaxies at redshifts z> 3},
  author={Steidel, Charles C and Giavalisco, Mauro and Pettini, Max and Dickinson, Mark and Adelberger, Kurt L},
  journal={The Astrophysical Journal},
  volume={462},
  number={1},
  pages={L17},
  year={1996},
  publisher={IOP Publishing}
}

@ARTICLE{sheth_tormen_1999_bias,
       author = {{Sheth}, Ravi K. and {Tormen}, Giuseppe},
        title = "{Large-scale bias and the peak background split}",
      journal = {\mnras},
         year = 1999,
        month = sep,
       volume = {308},
       number = {1},
        pages = {119-126},
          doi = {10.1046/j.1365-8711.1999.02692.x},
archivePrefix = {arXiv},
       eprint = {astro-ph/9901122},
 primaryClass = {astro-ph},
       adsurl = {https://ui.adsabs.harvard.edu/abs/1999MNRAS.308..119S}
}

@ARTICLE{smith+2003,
       author = {{Smith}, R.~E. and {Peacock}, J.~A. and {Jenkins}, A. and {White}, S.~D.~M. and {Frenk}, C.~S. and {Pearce}, F.~R. and {Thomas}, P.~A. and {Efstathiou}, G. and {Couchman}, H.~M.~P.},
        title = "{Stable clustering, the halo model and non-linear cosmological power spectra}",
      journal = {\mnras},
         year = 2003,
        month = jun,
       volume = {341},
       number = {4},
        pages = {1311-1332},
          doi = {10.1046/j.1365-8711.2003.06503.x},
archivePrefix = {arXiv},
       eprint = {astro-ph/0207664},
 primaryClass = {astro-ph},
       adsurl = {https://ui.adsabs.harvard.edu/abs/2003MNRAS.341.1311S}
}

@ARTICLE{smith+2007,
       author = {{Smith}, Robert E. and {Scoccimarro}, Rom{\'a}n and {Sheth}, Ravi K.},
        title = "{Scale dependence of halo and galaxy bias: Effects in real space}",
      journal = {\prd},
         year = 2007,
        month = mar,
       volume = {75},
       number = {6},
          eid = {063512},
        pages = {063512},
          doi = {10.1103/PhysRevD.75.063512},
archivePrefix = {arXiv},
       eprint = {astro-ph/0609547},
 primaryClass = {astro-ph},
       adsurl = {https://ui.adsabs.harvard.edu/abs/2007PhRvD..75f3512S}
}

@ARTICLE{smith+2011_nl_clustering,
       author = {{Smith}, Robert E. and {Desjacques}, Vincent and {Marian}, Laura},
        title = "{Nonlinear clustering in models with primordial non-Gaussianity: The halo model approach}",
      journal = {\prd},
         year = 2011,
        month = feb,
       volume = {83},
       number = {4},
          eid = {043526},
        pages = {043526},
          doi = {10.1103/PhysRevD.83.043526},
archivePrefix = {arXiv},
       eprint = {1009.5085},
 primaryClass = {astro-ph.CO},
       adsurl = {https://ui.adsabs.harvard.edu/abs/2011PhRvD..83d3526S}
}

@ARTICLE{scoccimarro_2001,
       author = {{Scoccimarro}, Rom{\'a}n and {Sheth}, Ravi K. and {Hui}, Lam and {Jain}, Bhuvnesh},
        title = "{How Many Galaxies Fit in a Halo? Constraints on Galaxy Formation Efficiency from Spatial Clustering}",
      journal = {\apj},
         year = 2001,
        month = jan,
       volume = {546},
       number = {1},
        pages = {20-34},
          doi = {10.1086/318261},
archivePrefix = {arXiv},
       eprint = {astro-ph/0006319},
 primaryClass = {astro-ph},
       adsurl = {https://ui.adsabs.harvard.edu/abs/2001ApJ...546...20S}
}

@ARTICLE{seljak_2000,
       author = {{Seljak}, Uro{\v{s}}},
        title = "{Analytic model for galaxy and dark matter clustering}",
      journal = {\mnras},
         year = 2000,
        month = oct,
       volume = {318},
       number = {1},
        pages = {203-213},
          doi = {10.1046/j.1365-8711.2000.03715.x},
archivePrefix = {arXiv},
       eprint = {astro-ph/0001493},
 primaryClass = {astro-ph},
       adsurl = {https://ui.adsabs.harvard.edu/abs/2000MNRAS.318..203S}
}

@article{scannapieco+2002_nlbias,
    author = "Scannapieco, Evan and Barkana, Rennan",
    title = "{An analytical approach to inhomogeneous structure formation}",
    eprint = "astro-ph/0205276",
    archivePrefix = "arXiv",
    doi = "10.1086/340063",
    journal = "Astrophys. J.",
    volume = "571",
    pages = "585",
    year = "2002"
}

@ARTICLE{schneider+2012_wdm,
       author = {{Schneider}, Aurel and {Smith}, Robert E. and {Macci{\`o}}, Andrea V. and {Moore}, Ben},
        title = "{Non-linear evolution of cosmological structures in warm dark matter models}",
      journal = {\mnras},
         year = 2012,
        month = jul,
       volume = {424},
       number = {1},
        pages = {684-698},
          doi = {10.1111/j.1365-2966.2012.21252.x},
archivePrefix = {arXiv},
       eprint = {1112.0330},
 primaryClass = {astro-ph.CO},
       adsurl = {https://ui.adsabs.harvard.edu/abs/2012MNRAS.424..684S}
}

@article{scranton2002analysis,
  title={Analysis of systematic effects and statistical uncertainties in angular clustering of galaxies from early Sloan Digital Sky Survey data},
  author={Scranton, Ryan and Johnston, David and Dodelson, Scott and Frieman, Joshua A and Connolly, Andy and Eisenstein, Daniel J and Gunn, James E and Hui, Lam and Jain, Bhuvnesh and Kent, Stephen and others},
  journal={The Astrophysical Journal},
  volume={579},
  number={1},
  pages={48},
  year={2002},
  publisher={IOP Publishing}
}

@ARTICLE{shuntov+2025,
       author = {{Shuntov}, Marko and {Oesch}, Pascal A. and {Toft}, Sune and {Meyer}, Romain A. and {Covelo-Paz}, Alba and {Paquereau}, Louise and {Bouwens}, Rychard and {Brammer}, Gabriel and {Gelli}, Viola and {Giovinazzo}, Emma and {Herard-Demanche}, Thomas and {Illingworth}, Garth D. and {Mason}, Charlotte and {Naidu}, Rohan P. and {Weibel}, Andrea and {Xiao}, Mengyuan},
        title = "{Constraints on the early Universe star formation efficiency from galaxy clustering and halo modeling of H$\alpha$ and [O III] emitters}",
      journal = {arXiv e-prints},
         year = 2025,
        month = mar,
          eid = {arXiv:2503.14280},
        pages = {arXiv:2503.14280},
          doi = {10.48550/arXiv.2503.14280},
archivePrefix = {arXiv},
       eprint = {2503.14280},
 primaryClass = {astro-ph.GA},
       adsurl = {https://ui.adsabs.harvard.edu/abs/2025arXiv250314280S}
}

@ARTICLE{tinker+2012_clustering_cosmology,
       author = {{Tinker}, Jeremy L. and {Sheldon}, Erin S. and {Wechsler}, Risa H. and {Becker}, Matthew R. and {Rozo}, Eduardo and {Zu}, Ying and {Weinberg}, David H. and {Zehavi}, Idit and {Blanton}, Michael R. and {Busha}, Michael T. and {Koester}, Benjamin P.},
        title = "{Cosmological Constraints from Galaxy Clustering and the Mass-to-number Ratio of Galaxy Clusters}",
      journal = {\apj},
         year = 2012,
        month = jan,
       volume = {745},
       number = {1},
          eid = {16},
        pages = {16},
          doi = {10.1088/0004-637X/745/1/16},
archivePrefix = {arXiv},
       eprint = {1104.1635},
 primaryClass = {astro-ph.CO},
       adsurl = {https://ui.adsabs.harvard.edu/abs/2012ApJ...745...16T}
}

@ARTICLE{Tinker+2010_bias,
       author = {{Tinker}, Jeremy L. and {Robertson}, Brant E. and {Kravtsov}, Andrey V. and {Klypin}, Anatoly and {Warren}, Michael S. and {Yepes}, Gustavo and {Gottl{\"o}ber}, Stefan},
        title = "{The Large-scale Bias of Dark Matter Halos: Numerical Calibration and Model Tests}",
      journal = {\apj},
         year = 2010,
        month = dec,
       volume = {724},
       number = {2},
        pages = {878-886},
          doi = {10.1088/0004-637X/724/2/878},
archivePrefix = {arXiv},
       eprint = {1001.3162},
 primaryClass = {astro-ph.CO},
       adsurl = {https://ui.adsabs.harvard.edu/abs/2010ApJ...724..878T}
}

@article{Tinker_2005_bias,
doi = {10.1086/432084},
url = {https://dx.doi.org/10.1086/432084},
year = {2005},
month = {sep},
publisher = {},
volume = {631},
number = {1},
pages = {41},
author = {Tinker, Jeremy L. and Weinberg, David H. and Zheng, Zheng and Zehavi, Idit},
title = {On the Mass-to-Light Ratio of Large-Scale Structure},
journal = {The Astrophysical Journal}
}

@ARTICLE{toshikawa+2024,
       author = {{Toshikawa}, Jun and {Wuyts}, Stijn and {Kashikawa}, Nobunari and {Liu}, Chengze and {Sawicki}, Marcin and {Overzier}, Roderik and {Kubo}, Mariko and {Uchiyama}, Hisakazu and {Ito}, Kei and {Bremer}, Malcolm and {Ono}, Yoshiaki and {Kodama}, Tadayuki and {Lin}, Yen-Ting and {Saito}, Tomoki},
        title = "{An enhanced abundance of bright galaxies in protocluster candidates at z   3-5}",
      journal = {\mnras},
         year = 2024,
        month = jan,
       volume = {527},
       number = {3},
        pages = {6276-6291},
          doi = {10.1093/mnras/stad3162},
archivePrefix = {arXiv},
       eprint = {2310.08525},
 primaryClass = {astro-ph.GA},
       adsurl = {https://ui.adsabs.harvard.edu/abs/2024MNRAS.527.6276T}
}

@article{toshikawa2018goldrush,
  title={GOLDRUSH. III. A systematic search for protoclusters at z~ 4 based on the> 100 deg2 area},
  author={Toshikawa, Jun and Uchiyama, Hisakazu and Kashikawa, Nobunari and Ouchi, Masami and Overzier, Roderik and Ono, Yoshiaki and Harikane, Yuichi and Ishikawa, Shogo and Kodama, Tadayuki and Matsuda, Yuichi and others},
  journal={Publications of the Astronomical Society of Japan},
  volume={70},
  number={SP1},
  pages={S12},
  year={2018},
  publisher={Oxford University Press}
}

@article{tanaka2018photometric,
  title={Photometric redshifts for Hyper Suprime-Cam Subaru strategic program data release 1},
  author={Tanaka, Masayuki and Coupon, Jean and Hsieh, Bau-Ching and Mineo, Sogo and Nishizawa, Atsushi J and Speagle, Joshua and Furusawa, Hisanori and Miyazaki, Satoshi and Murayama, Hitoshi},
  journal={Publications of the Astronomical Society of Japan},
  volume={70},
  number={SP1},
  pages={S9},
  year={2018},
  publisher={Oxford University Press}
}

\bsp	
\label{lastpage}
\end{document}